\documentclass[10pt]{IEEEtran}

\usepackage{color}
\usepackage{placeins}
\usepackage{float}
\usepackage[hidelinks]{hyperref}
\usepackage{tabularx,colortbl}
\usepackage[cmex10]{amsmath}
\usepackage{physics}
\usepackage{amssymb}
\usepackage[dvipsnames]{xcolor}

\usepackage[pdftex]{graphicx}
\graphicspath{{./Figures/}}
\usepackage{subcaption}
\usepackage{booktabs}
\usepackage{bm}

\def\d{\mathrm{d}}
\DeclareMathOperator{\sgn}{sgn}

\mathchardef\mhyphen="2D

\begin{document}
%
\title{Near-Field Angular Scan Enhancement of Antenna Arrays Using Metasurfaces}


\author{Gleb~A.~Egorov, and~George~V.~Eleftheriades,~\IEEEmembership{Fellow,~IEEE}
\thanks{G. A. Egorov is with the Edward S. Rogers Department of Electrical and Computer Engineering, University of Toronto,
Toronto, Ontario, Canada (e-mail: gleb.egorov@mail.utoronto.ca).}
\thanks{G. V. Eleftheriades is with the Edward S. Rogers Department of Electrical and Computer Engineering, University of Toronto,
Toronto, Ontario, Canada (e-mail: gelefth@waves.utoronto.ca).}}


%


\maketitle

\begin{abstract}
In this publication we continue our previous work on extending the angular scan range of phased arrays using metasurfaces. We consider in detail scan enhancement using a single metasurface lens in the near-field of a source and provide analytical expressions for source excitation to obtain a desired beam. We show that such a device suffers from the same directivity degradation as in the case of far-field lens placement. Moreover, we discuss in detail that this approach has limitations, and one cannot place the lens arbitrarily close to the source array. We then propose a two-lens near-field scan enhancer which is not subject to the same limitations, although still subject to the same directivity degradation. Finally we propose a binary metasurface concept which theoretically achieves desired scan enhancement without directivity degradation. Some theoretical claims are then verified via full-field simulations.
\end{abstract}

\begin{IEEEkeywords}
Phased arrays, beam steering, lenses, metasurfaces.
\end{IEEEkeywords}

%
\IEEEpeerreviewmaketitle

\section{Introduction}

In this paper we contniue our work on angular scan enhancement of phased antenna arrays using metasurface techniques. This paper is a direct continuation of our previous work \cite{egorov2020theory}, of which Sections II and IV are of relevance. We continue by focusing our attention to scan-enhancing devices which are placed in the radiating near-field of a source phased array. 

The problem of angular scan enhancement can be introduced as follows. In communication systems employing phased arrays (such as 5G), it is desired to obtain maximum possible directivity from an array while having it be able to scan up to some desired $\theta_{max}$ off-broadside. This can be achieved in two simple ways -- by increasing the number of antenna elements in the array, or increasing the spacing between the elements. Both approaches lead to an increased effective aperture size of the antenna, thus increasing the directivity. Increasing the number of elements is often undesired, since the complexity of the array feed network and cost increase non-linearly with the number of elements. Increasing element spacing works well up to $\lambda/2$ spacing. Beyond $\lambda/2$ spacing, the array is capable of producing a grating lobe and its scan may be compromised. For conventional arrays, spacing beyond $\lambda$ is undesired since the array produces a grating lobe no matter the phasing. Some exotic array architectures are able to obtain a limited grating-lobe-free scan range with effective element spacing even larger than $\lambda$. Such array architectures can be obtained via clustering, random sub-arraying, overlapping, and interleaving \cite{rocca2016unconventional,azar2013overlapped}. Imagine we choose a large enough spacing, larger than $\lambda/2$, such that the array is not capable of steering to $\theta_{max}$ due to the presence of a grating lobe. Is it possible to enhance the grating-lobe-free angular scan range of the array to the desired $\theta_{max}$ using some metasurface, or otherwise, device placed above the array? This question has been considered in a number of publications, spanning various types of devices, which we summarize below.

Reference \cite{steyskal1979gain} is one of the earliest theoretical investigations, where the authors considered a hypothetical infinitesimally thin phase-incurring dome placed above an array, with a focus on obtaining directive radiation all the way to the horizon. In \cite{kawahara2007design} a physical dielectric dome was fabricated and tested. Scan enhancing devices designed via transformation optics techniques were analyzed in \cite{lam2011steering,moccia2017transformation}. Power conservation arguments were used in \cite{kazim2013advanced,kazim2016wide} to obtain enhancement via shaping of gain pattern envelope of a source array. These publications share some similarities. Regardless of the design approach, all proposed devices resemble simple diverging lenses. Although the effect of the scan-enhancing devices on directivity has not been analyzed theoretically, all simulated and measured data show that a scan enhancement appears to always be accompanied by a degradation in directivity. 

Perhaps the most promising scan enhancement approach up to date is discussed in \cite{benini2018phase}. There, a hypothetical phase-incurring lens-like surface was used to achieve scan enhancement. To achieve directive radiation, the source array was excited with non-uniform magnitude and non-linear phase as dictated by time-reversed numerical arguments. In this way, the metasurface radome could be placed in close proximity to the phased array. Nevertheless, we show that this method also leads to a directivity degradation combined with a possible appearance of a ``distributed" grating lobe which can further degrade performance. 

To arrive at the above conclusions we first define ideal refraction and non-paraxial ray theory in Sec. \ref{sec:Prelim}. These definitions allow us to study the problem of near-field scan enhancement using a single metasurface lens analytically, instead of relying on numerical expressions as in \cite{benini2018phase}. In Sec. \ref{sec:SingleLensTheory} we solve the single lens enhancer for the required array excitation to obtain directive radiation in a desired direction. In Sec. \ref{sec:LimSingleLens} we show that the near-field single-lens enhancer not only leads to a directivity degradation, but is subject to another limitation which dictates how close the lens can be placed to the source array. To circumvent the limitation of the single-lens enhancer, we propose a two-lens enhancing device in Sec. \ref{sec:TwoLensTheory}. We show that  the two-lens device is still subject to the same directivity degradation. In Sec. \ref{sec:Tunable}, which concludes the theoretical considerations, we propose a binary (two-state) tunable metasurface concept which would achieve a desired near-field scan enhancement with no degradation in directivity. 

Select theory of Sec. \ref{sec:Theory} is put to the test in Sec. \ref{sec:NumStudies}. Section \ref{sec:DGLCond} is a detailed comparison of derived limitations of the single-lens enhancer, but which also discusses when the presented theory can be considered accurate or not. In Sec. \ref{sec:SingleLensSims} we simulate various single-lens scan enhancers and observe good agreement between theory and simulation. Finally, Sec. \ref{sec:TwoLensSims} shows simulation results of various two-lens enhancers and a comparison is made between a single- and two-lens device. Again, close agreement between theory and simulation is observed. 

Note that most of the devices considered in this publication are composed of lenses (in the sense of the well-known lens phase, see (\ref{eq:LensPhase})) to achieve a desired scan enhancement. Nevertheless these lenses have to be considered as metasurfaces because they can be achieved physically only using metasurface techniques. The considered frequency and size of the problem leads us to consider extremely strong lenses, ones with focal lengths on the order of a few wavelengths. Common dielectric lenses cannot in general achieve such performance, and if they can oftentimes the lenses have to be extremely thick. Thus when we refer in this publication to lenses, we imply metasurface lenses. 

Throughout this publication we assume $\exp(j\omega t)$ dependence. All geometries and thus fields have two-dimensional dependence. The problem is considered to be two-dimensional due to our inability to simulate three-dimensional scenarios. All considered sources are composed of out-of-plane infinite lines of electric current. This results in us considering TE fields only. Similar arguments can be made for the TM polarization. COMSOL Multiphysics in conjunction with MATLAB were used to obtain simulated data. The more mathematical aspects were evaluated with Wolfram Mathematica.

\section{Theory} \label{sec:Theory}

\subsection{Preliminary Considerations} \label{sec:Prelim}

\begin{figure}[t!]
\begin{center}
\includegraphics[width=3.5in]{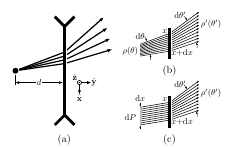}
\caption{(a) A single lens far-field scan enhancer is shown along with coordinate system vectors used throughout this publication. (b) Depiction of ideal refraction for the far-field case. The source array can be considered as a point source since the lens is placed in the far-field. (c) Depiction of ideal refraction for the incident near-field case. Note that it can be thought as if a single ray in (b) and (c) carries a nominal amount of power. The ideal refraction statements of (\ref{eq:IdealRefFF}) and (\ref{eq:IdealRefdN}) are implied in the diagrams via the fact that the number of rays remains constant upon passing through the lens. }\label{fig:PrelimCons}
\end{center}
\end{figure}

In \cite{egorov2020theory} we showed that placing a diverging lens in the far-field of an antenna array can achieve angular scan enhancement of array beams. Assuming the array produces a beam at an angle $\theta$ off-broadside, the lens will refract the beam to point in the direction $\theta'=\alpha\theta$, where $\alpha$ was termed the scan enhancement factor. Given a desired $\alpha$, the array-lens distance and the focal length of the lens are not independent, but are related according to
\begin{equation} \label{eq:dfalpha}
\frac{d}{f}=1-\alpha.
\end{equation}
The single lens scan enhaner is depicted in Fig. \ref{fig:PrelimCons}(a). In this publication we are interested in placing scan-enhancing lenses in the radiating near-field of array sources. Although $\alpha$ was rigorously discussed only for the case of far-field lens placement, we will continue to use the same terminology to imply the $d/f$ relationship of (\ref{eq:dfalpha}). Furthermore, we are interested in angular scan enhancement without beam angle inversion ($\alpha > 1$), and so only diverging lenses are considered. Thus, all focal lengths $f$ appearing below are assumed to be negative, except for when an expression explicitly contains ``$\sgn(f)$" as in (\ref{eq:LensPhase}). Note that identical reasoning can be applied to converging lenses (positive focal lengths) and similar results and expressions can be easily obtained. A single converging lens can achieve $\alpha < 1$ -- angular scan enhancement with beam direction inversion for $\alpha < -1$ and scan degradation with ($-1 < \alpha < 0$) and without ($0 \le \alpha < 1$) direction inversion. On a related note, according to \cite{egorov2020theory}, scan degradation is accompanied by directivity enhancement, and so one can discuss ``directivity enhancement" of antenna arrays using a converging lens.

When discussing theoretical aspects of scan enhancement we assume the considered lenses to refract ideally. Ideal refraction implies refraction with energy conservation and no reflections. In the case when a lens is placed in the far-field of a source, ideal refraction is described as follows (see Fig. \ref{fig:PrelimCons}(b) for reference). The radiation of the source is described by the ray density function $\rho(\theta)$ \cite{egorov2020theory}. Consider the rays which are incident on an infinitesimal section of the lens $[x, x+\d x]$. These rays happen to span an angular range given by $[\theta,\theta+\d\theta]$. The lens refracts these rays into the angular range $[\theta',\theta'+\d\theta']$ according to the laws of geometric optics. Ideal refraction is then defined as
\begin{equation} \label{eq:IdealRefFF}
\rho(\theta)\d\theta=\rho'(\theta')\d\theta',
\end{equation}
where $\rho'(\theta')$ is the ray density of the refracted beam. This statement of ideal refraction was used in \cite{egorov2020theory} when discussing the effects of scan-enhancing lenses on the directivity of incident beams. 

Another case where we assume ideal refraction occurs is when incident radiation is collimated and the refraction produces far-fields (see Fig. \ref{fig:PrelimCons}(c) for reference). In this case, the incident radiation is not in the far-field and no $\rho(\theta)$ can be attributed. Since the refracted fields are assumed to be far-field (note that these fields do not have a ``physical" source, and originate from a virtual image point behind the lens as will be discussed later on), there is a $\rho'(\theta')$. The collimated beam has a total power $P$, and $\d P$ is incident on the lens in the range $[x, x+\d x]$. In this range, the radiation is refracted into $[\theta',\theta'+\d\theta']$ directions. For this case, ideal refraction is defined as
\begin{equation} \label{eq:IdealRefdN}
\d P=\rho'(\theta')\d\theta'.
\end{equation}

The final consideration regards extending the ray transfer matrix theory (see \cite{yariv2007photonics} for description of the theory) to non-paraxial rays. This paper studies lenses for scan enhancement, thus we are only interested in matrices which describe ray propagation through free space and ideal lenses. In the context of the standard theory, the state of a ray at a given position along the optic axis is described by a two element vector -- giving the distance $x$ of the ray to the optic axis and the angle $\theta$ the ray makes with respect to the optic axis. In order for the same two matrices to describe propagation of non-paraxial rays through free space and ideal lenses, the state vectors have to be transformed as 
\begin{equation}
\begin{bmatrix}
x \\
\theta
\end{bmatrix} 
\to
\begin{bmatrix}
x \\
\tan\theta
\end{bmatrix}.
\end{equation}
It is easy to check that the new state vectors when used with the two matrices of interest lead to consistent behavior with respect to canonical rays. Canonical rays are the trivial rays of a single lens -- ones going through/aimed at one of the focal points and the rays passing through the center of the lens. Note that our proposed non-paraxial ray optics is in general inconsistent with the more physically fundamental generalized law of refraction. Appendix \ref{sec:AppNonParaRerf} discusses when the proposed treatment of refraction can be considered approximately accurate.

\subsection{Near-Field Scan Enhancement using a Single Lens} \label{sec:SingleLensTheory}

In \cite{egorov2020theory} we discussed the possibility of placing scan-enhancing lenses in the radiating near-field of source arrays. However, we have done so only for the case of an angle-doubling setup ($\alpha=2$) and broadside incidence. Furthermore, we did not discuss in detail how the array is excited to obtain directive radiation after lens refraction. In order to obtain good performance one cannot keep exctiting the array using linear phase and uniform magnitude. In \cite{egorov2020theory} we showed that if such array excitation is applied, the directivity degrades indefinetely as the scan enhancing lens is brought closer. In this section we discuss the required array excitations and generalize the results of \cite{egorov2020theory}. Note that the array excitation approach presented here is related to the method discussed in \cite{benini2018phase}, but we are able to provide analytical expressions instead of purely numerical ones.

\begin{figure}[t!]
\begin{center}
\includegraphics[width=3.5in]{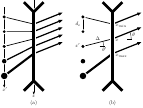}
\caption{The near-field single lens scan enhancer. (a) Depicts the source array (composed of infinite current lines extending in/out of the page), the $x$ and $x'$ axes and the rays producing directive radiation in some desired direction $\theta$. Ray thicknesses represent the power carried by the rays. Refracted rays are uniform, implying a uniform aperture on the output. The source array has to be excited such that a uniform aperture is established. The incident ray thicknesses follow (\ref{eq:rhoprime}) qualitatively. The time-forward case is shown, and the time-reversed scenario is obtained simply by reversing ray directions. (b) Shows some of the defined quanitites used in the analysis. One can discuss a particular ray simply by refering to either its $x$, $x'$ or $\theta'$ values since they are not independent.}\label{fig:SingleLensNearField}
\end{center}
\end{figure}

Let us define the geometry of the problem as follows (see Fig. \ref{fig:SingleLensNearField} for reference). A right-hand coordinate system is assumed. The lens lies along the $x$-axis. The optic axis of the lens is coincident with the $y$-axis. Directions $\hat{x}$ and $\hat{y}$ are chosen to be consistent with the coordinate system of \cite{egorov2020theory}. A linear array, composed of infinite current lines extending in the $\hat{z}$ direction, is placed some distance $d$ behind the lens. The geometrical center of the array lies on the optic axis and the array extends in the $\hat{x}$ direction. The array exhibits an inter-element distance $d_e$. We consider the case when $\lambda/2 < d_e < \lambda$. This condition guarantees that the linearly phased array is capable of producing a single (and not more) grating lobe. 

The unknown antenna array excitation is composed of two parts -- the phasing of the elements and the magnitude with which each element is excited. And so, let us consider these two parts of the overall excitation separately. 

\subsubsection{Antenna Element Phasing for Constructive Interference in a Desired $\theta$}

We are interested in producing directive radiation with the array-lens setup far away at some angle $\theta$ off-broadside (i.e. away from the optic axis) as shown in Fig. \ref{fig:SingleLensNearField}. Each array element on its own produces rays which emanate in all directions. Each ray can be attributed to some Fourier coefficient of the produced field, i.e. a plane wave \cite{landau2013classical}. From this analogy, we can say that as a ray propagates from an element, the field associated with it acquires a phase of $-kl$ (negative due to the assumed $\exp(j\omega t)$ dependence), with $l$ being the distance from the element to the point of interest along the ray. In the presence of a few radiating elements, the produced field at some point of interest depends on the interference of the rays from each element which pass through the point. And so, we want all the element rays to interfere constructively at a point infinitely far away at an angle $\theta$. Thus, we are interested in rays which travel in direction $\theta$ after being refracted by the lens. We assume that the path of a ray passing through the lens is dictated by non-paraxial ray transfer matrix theory. Each element produces a single ray which exits the lens at $\theta$. Because rays have a phase attributed to them, we can ask whether a ray would acquire any phase while passing through an ideal lens. 

Consider the case of parallel, coherent rays being incident on a diverging lens. These rays will refract as if they were emitted from a point source at the focal point behind the lens. Thus, the phase of refracted rays along the transmission side of the lens is $-k\sqrt{x^2+f^2}$. On the incident side of the lens the rays are in phase. Thus we can say that as a ray passes through the lens it acquires a phase of
\begin{equation} \label{eq:LensPhase}
\phi(x) = \sgn(f)k\sqrt{x^2+f^2}.
\end{equation}
Note that this expression is valid for focusing and diverging lenses. We now assume that this phase function remains valid for all possible incidence scenarios. 

The condition that all the rays of interest interfere constructively infinitely far away is equivalent to the refracted rays of interest having the same phase along any plane perpendicular to the direction $\theta$. One of these rays originates from an array element at some position $(x',-d)$ and propagates in direction $\theta'$ (see Fig. \ref{fig:SingleLensNearField}(b)). This ray arrives at the lens at some position $(x,0)$ and is refracted to $\theta$. The antenna element which produced this ray did so with some initial phase $\phi_e(x')$. As the ray propagates towards the lens it covers a distance $\Delta$ (which can be expressed as a function of $x'$ or $x$ or a combination of the two) and acquires the phase of $-k\Delta$. Once the ray passes through the lens at $x$, it acquires the phase given by (\ref{eq:LensPhase}). The ray now travels the distance $x \sin\theta$ to a plane perpendicular to the direction $\theta$ and acquires the phase $-kx\sin\theta$. At this point it is desired that all the rays of interest reach this perpendicular plane with the same phase. This can be achieved by choosing the array element phase according to
\begin{equation} \label{eq:GeneralPhasing}
\phi_e(x')=k\Delta(x,x')-\phi(x)+kx\sin\theta.
\end{equation}
It is easy to see that 
\begin{equation} \label{eq:Delta}
\Delta(x,x') = \sqrt{d^2+(x-x')^2}.
\end{equation}
We can now employ non-paraxial ray transfer theory to express (\ref{eq:GeneralPhasing}) in terms of $x'$ only. 

In terms of non-paraxial ray transfer theory, the path of a ray from an antenna element through the lens is described by \cite{yariv2007photonics}
\begin{equation} \label{eq:RayMatrix}
\begin{bmatrix}
-x \\
\tan\theta
\end{bmatrix} 
=
\begin{bmatrix}
1 & d \\
-\frac{1}{f} & \alpha
\end{bmatrix} 
\begin{bmatrix}
-x' \\
\tan\theta'
\end{bmatrix}.
\end{equation}
The negative signs appearing before $x$ and $x'$ are there because the $\hat{x}$ we have chosen is opposite to the convention of standard ray transfer theory. Eq. (\ref{eq:RayMatrix}) is easily rearranged to obtain
\begin{align} \label{eq:ABCDrelation1}
\alpha x &= x'-d\tan\theta, \\ 
\alpha f\tan\theta' &= -x'+f\tan\theta. \label{eq:ABCDrelation2}
\end{align}
Using (\ref{eq:LensPhase}), (\ref{eq:Delta}) and (\ref{eq:ABCDrelation1}) in (\ref{eq:GeneralPhasing}) we obtain
\begin{multline} \label{eq:PhasingRayTheory}
\phi_e(x') = \frac{kd}{\alpha |f|}\sqrt{\alpha^2f^2+\left(x'-f\tan\theta\right)^2}+\\ \frac{k}{\alpha}\sqrt{\alpha^2f^2+\left(x'-d\tan\theta\right)^2}+\\ \frac{k\sin\theta}{\alpha}\left(x'-d\tan\theta\right).
\end{multline}
The above equation is the analytical expression for the required element phasing we were after.

\subsubsection{Antenna Element Current Magnitudes for a Uniform Output Aperture}

We would like to obtain the largest possible directivity in the direction $\theta$. So far element phasing was considered to achieve constructive interference. Now we focus on the magnitude of element excitation which would lead to maximum directivity. It is known that peak directivity of a uniform aperture is given by \cite{egorov2020theory}
\begin{equation} \label{eq:DUmax}
D_{U,max}=\frac{2\pi L}{\lambda}.
\end{equation}
Furthermore, it is known that given an aperture of a certain length the aperture achieves its maximum possible directivity when it is uniform (excluding the super-directivity case) \cite{silver1984microwave}.
For our problem, the size of the output aperture is dictated by the $x$ locations of the rays of interest during refraction. Once the array is phased such that constructive interference occurs in the desired $\theta$, the size of the output aperture is set. The directivity towards $\theta$ would be the maximum possible value if the output aperture were uniform. Thus, we would like to excite the array elements with magnitudes which would approximately produce a uniform aperture on the transmission side of the lens. 

Time reversal can be applied to this scnario \cite{benini2018phase}. And so we consider the reverse situation -- a uniform, collimated beam is incident on the transmission side of the lens at an angle $\theta$. The beam has a total power $P$ and illuminates a section of the lens of length $L$. The lens diverges this beam and we assume far-fields are set up on the incident side of the lens. This assumption allows us to use ideal refraction in the sense of (\ref{eq:IdealRefdN}). At this point we want to calculate the far-field ray density $\rho'(\theta')$ created by the lens. From (\ref{eq:IdealRefdN}), $\rho'(\theta')=\d P/ \d \theta'$. Due to the uniformity of the incident beam, $\d P=\frac{P}{L}\d x$. This leads to
\begin{equation}
\rho'(\theta')=\frac{P}{L} \left| \frac{\d x}{\d \theta'} \right|.
\end{equation} 
The absolute value of the derivative ensures the ray density has physical meaning (i.e. is positive). Combining (\ref{eq:ABCDrelation1}) and (\ref{eq:ABCDrelation2}) gives
\begin{equation} \label{eq:ABCDrelation3}
\alpha f \tan\theta' = -\alpha x + (f-d)\tan\theta.
\end{equation}
Performing implicit differentiation with respect to $\theta'$ leads to
\begin{equation} \label{eq:rhoprime}
\rho'(\theta') = \frac{P |f|}{L} \sec^2 \theta'.
\end{equation}
Note that the above is valid for the rays appearing in the $y<0$ region of Fig. \ref{fig:SingleLensNearField}(a) (with time reversal in mind).

Note that from the point of view of non-paraxial ray optics, the assumption that a far-field $\rho'(\theta')$ is established is perfectly valid. Using (\ref{eq:ABCDrelation3}) it is possible to trace the refracted rays behind the lens, and it is easy to see that these rays all intersect at a single virtual image point $(f\tan\theta,-f)$. Thus, it appears that $\rho'(\theta')$ is produced by a source of infinitesimal size in terms of wavelengths. For such sources, far-fields are established at any finite distance away (without accounting for the near-field region of the cylindrical sources themselves, which is approximately $\lambda/4$). In reality, refraction is dictated by the generalized refraction law (see Appendix \ref{sec:AppNonParaRerf}), which would lead to a finitely sized virtual image. However, as long as non-paraxial ray optics remains approximately valid, we expect the size of the virtual image to remain small and still lead to a far-field $\rho'(\theta')$. The next natural question to consider with regard to the validity of the far-field assumption is whether the far-fields fields implied by (\ref{eq:rhoprime}) are consistent with Maxwell's equations. This is discussed in detail in Appendix \ref{sec:AppRhoPrimeValid}, and it turns out that as long as $|f|$ is approximately larger than $1.6\lambda$, the far-fields can be considered approximately valid. 

Using (\ref{eq:rhoprime}), the electric far-field produced by the virtual source point can be written as \cite{egorov2020theory, vladimirov1976equations, harrington1961time}
\begin{equation} \label{eq:Eff}
\mathbf{E}(r',\theta') = \hat{\mathbf{z}} \hphantom{,} \sqrt{\frac{\rho'(\theta')}{r'}} \hphantom{,} e^{-jk r'},
\end{equation}
where $r'$ is the distance from the virtual source point to some observation point. Recall from above, that in the far-field rays correspond to plane waves, and thus we can say that the far-field magnetic field associated with (\ref{eq:Eff}) is
\begin{equation} \label{eq:Hff}
\mathbf{H}(r',\theta') = \frac{1}{\eta} \hphantom{,} \sqrt{\frac{\rho'(\theta')}{r'}} \hphantom{,} e^{-jk r'} 
\begin{bmatrix}
\cos\theta' \\
\sin\theta' \\
0
\end{bmatrix}.
\end{equation}
We now apply Love's equivalence theorem \cite{harrington1961time}, which allows us to place electric and magnetic surface currents along the $y=-d$ plane such that the fields in the $y>-d$ region are given by (\ref{eq:Eff}) and (\ref{eq:Hff}) and in the region $y<-d$ the fields are 0. All together, this corresponds to the time-reversed beam being refracted by the lens, and subsequently absorbed by the currents. It is easy to see that in the time-forward case these currents would produce the desired beam in direction $\theta$. Because these currents produce plane waves, we can set the magnetic current to 0 without affecting the desired forward beam (apart from overall scaling). Disregarding the magnetic current leads to two produced ``beams" in opposite directions. This is obvious from symmetry considerations of radiated fields from the electric currents. This surface electric current is the ideal current sheet which produces the desired uniform beam in $\theta$. 

The source of the problem is an antenna array, not a continuous current sheet. Thus, we have to try to approximate (in some sense) the ideal current sheet by the discrete cylindrical sources of the array. There is no unique way to do this, and we provide the most trivial option -- direct sampling of the plane wave spectrum of the time-reversed scenario. With the summetry of the forward/backward beams produced by the electric current sheet in mind, the standard electric current boundary condition along $y=-d$ is written as \cite{harrington1961time}
\begin{equation}
\mathbf{J} = -\hat{\mathbf{z}} \hphantom{,} 2 \left(\hat{\mathbf{x}}\cdot\mathbf{H}(r',\theta')\right).
\end{equation}
This equation is further evaluated by plugging in (\ref{eq:rhoprime}) and (\ref{eq:Hff}). Note that we do not care for the phase of the ideal current sheet because we already know the required phasing of the antenna elements from (\ref{eq:PhasingRayTheory}). Furthermore, the overall scaling of the current sheet does not matter either because we are interested in the directivity performance, not how much total power is radiated. With these comments in mind, the ideal current sheet reduces to
\begin{equation}
|\mathbf{J}| = \sqrt{\frac{1}{r'}}.
\end{equation}
The distance between the virtual source point and an array element at $(x',-d)$ is
\begin{equation}
r' = \sqrt{(x'-f\tan\theta)^2 + (f-d)^2}.
\end{equation}
We choose the current of the antenna elements to be equal to the current sheet density evaluated at the element locations -- this in effect samples the plane waves which reach the elements in the time-reversed case. Thus, the required current magnitudes of the elements to achieve a uniform aperture in the time-forward case are given by
\begin{equation} \label{eq:Ie}
|I_e(x')| = \left( (x'-f\tan\theta)^2 + (f-d)^2 \right)^{-1/4},
\end{equation}
which is what we were after.

\subsection{Limitations of the Single Lens Enhancer} \label{sec:LimSingleLens}

Although the described above array excitation allows one to place a scan enhancing lens close to the source, the approach suffers from some drawbacks. First of all, the directivity is affected. This can be analyzed by comparing the output effective aperture length with the effective aperture length of the array itself at a given desired $\theta$. The effective aperture of the array when scanned to $\theta$ is $L_{arr}=L\cos\theta$. The output effective aperture length is given by
\begin{equation}
L_{eff} = (x_{max}-x_{min})\cos\theta,
\end{equation}
where $x_{max}= \frac{L}{2\alpha}-\frac{d}{\alpha}\tan\theta$ and $x_{min}= -\frac{L}{2\alpha}{-}\frac{d}{\alpha}\tan\theta$, which correspond to the paths of the outmost two rays of interest (see Fig. \ref{fig:SingleLensNearField}(b)). For small enough $\theta$ the array does not produce any grating lobes, and its peak directivity is described well by (\ref{eq:DUmax}). Simplifying $L_{eff}$ and using the aperture lengths in (\ref{eq:DUmax}) leads to a directivity degradation given by
\begin{equation} \label{eq:DirDegSingle}
\frac{D(\theta)}{D_{arr}(\theta)}=\frac{L_{eff}}{L_{arr}}=\frac{1}{\alpha},
\end{equation}
which is similar to the directivity degradation in the case of far-field lens placement with simple linear phasing of the elements \cite{egorov2020theory}.

Another drawback of the single lens enhancer is that there exists a limitation on how close one can place the lens to the source array. This occurs because as a lens is brought closer to the array, the non-linear inter-element phase increases, which in turn leads to power radiated in undesired directions. This behavior was discussed in \cite{egorov2020theory} for the case of $\theta=0$ and $\alpha=2$. Here we provide more general results. For a positive $\theta$, the two outermost elements at the $x'=L/2$ array end exhibit the maximum phase difference, given by
\begin{equation} \label{eq:deltaphimax}
\Delta\phi_{max} = \phi_e\left(\frac{L}{2}\right)-\phi_e\left(\frac{L}{2}-d_e\right).
\end{equation}
For a negative $\theta$, due to the symmetry of the setup with respect to the optic axis, the maximum phase difference occurs for the elements on the $x'=-L/2$ array end. Again due to symmetry, the absolute value of $\Delta\phi_{max}$ for a negative $\theta$ is the same as the value for a positive $\theta$ given by (\ref{eq:deltaphimax}). Furthermore, the whole problem exhibits even symmetry in the $x$-axis and anything of interest will have the same value at $-\theta$ as $\theta$. Thus considering only positive $\theta$ is enough to know overall performance, and we will do so for the rest of this section.

The two elements exhibiting maximum phase difference can be considered as a two-element array on its own. This two-element array will produce the peak of a grating lobe if the inter-element phase is larger than $2\pi-kd_e$. The appearance of the grating lobe peak of the two-element array signifies that power is being  lost to undesired directions which would reduce the overall directivity performance. As the lens is brought even closer, more elements will have excessively large inter-element phasing and will contribute to the power leakage, which we term as a distributed grating lobe. Thus we can say that if 
\begin{equation} \label{eq:GratingLobeCond}
\Delta\phi_{max} \ge 2\pi-kd_e,
\end{equation}
then the directivity performance of the device is compromised. Note that this reasoning does not provide an estimate on the directivity degradation, and is meant as a rule-of-thumb. Plugging in all the available expressions into the above inequality leads to a cumbersome expression with no simple interpretation. For the case of $\theta=0$, (\ref{eq:GratingLobeCond}) simplifies to
\begin{equation}
\sqrt{1+\frac{L^2}{4\alpha^2 f^2}} - \sqrt{1+\frac{(L-2d_e)^2}{4\alpha^2 f^2}}  \ge \frac{2\pi-kd_e}{k(d+|f|)}.
\end{equation}

A different reasoning approach leads to a much simpler (but less rigorous) condition for the creation of the distributed grating lobe. For a positive $\theta$, the most extreme angle $\theta'_{max}$ occurs for the outermost ray of interest originating from $x'=L/2$. According to (\ref{eq:ABCDrelation2}),
\begin{equation}
\theta'_{max} = \arctan \left( \frac{-L/2+f \tan\theta}{\alpha f} \right).
\end{equation} 
We expect that had the two outer elements been on their own they would produce a beam in some direction which is at most $\theta'_{max}$. Note that this statement does not have a logically sound argument and this is where the non-rigor of the reasoning comes from. It is known that a linearly phased array produces a grating lobe if it is steered to some angle larger than
\begin{equation}
\theta_g=\arcsin\left(\frac{\lambda}{d_e}-1\right).
\end{equation} 
Thus, if $\theta'_{max} \ge \theta_g$ then the array produces a distributed grating lobe and the directivity performance is compromised. This condition can be written as
\begin{equation} 
L \ge \frac{2|f|\alpha(\lambda-d_e)}{\sqrt{2\lambda d_e-\lambda^2}}-2 |f| \tan\theta.
\end{equation}
Using (\ref{eq:dfalpha}) we rewrite the above inequality as
\begin{equation}\label{eq:GratingLobeLlim}
\frac{L}{d} \ge \frac{2\alpha}{\alpha-1}\frac{1-\frac{d_e}{\lambda}}{\sqrt{2\frac{d_e}{\lambda}-1}}-\frac{2}{\alpha-1}\tan\theta.
\end{equation}
Although less logically sound, this inequality explicitly shows that having a too large of an array, or alternatively having the array too close to the lens, leads to distributed grating lobes. The two conditions (\ref{eq:GratingLobeCond}) and (\ref{eq:GratingLobeLlim}) are compared in Sec. \ref{sec:DGLCond}.

\subsection{Near-Field Scan Enhancement using Two Lenses} \label{sec:TwoLensTheory}

In the preceding subsections we built on the results of \cite{egorov2020theory} and showed that it is possible to bring a single lens scan enhancer into the radiating near-field of the source array while maintaining the same directivity performance as the far-field single lens enhancer. We then discussed that this close placement suffers from some limitations on how close one can place the lens to the array, how large of an array can be used and how large $\theta$ can be before power is lost to the distributed grating lobe. Here we discuss that using a two-lens scan enhancer in the near-field of the source array still leads to a similar directivity degradation as that of the far-field single lens case, while not suffering from the distributed grating lobe limitation. Moreover, no special weighting of the elements is required, just standard linear phasing. 

\begin{figure}[t!]
\begin{center}
\includegraphics[width=3.5in]{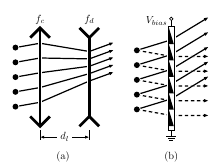}
\caption{(a) A two-lens scan enhancer, composed of a converging (focal length $f_c$) and a diverging lens (focal length $f_d$), angularly enhancing a beam produced by a linearly phased, unifromly excited antenna array. (b) Discretely tunable metasurface concept to achieve scan enhancement. $V_{bias}$ represents a required bias voltage which tunes the surface. The surface is drawn in a way which implies a linear, $2\pi$-wrapped phase. The surface is depicted in its first state ($\phi(x)=k_sx$), and prodives angular enhancement to the two shown incident beams. The dashed beam exits the device at broadside.}\label{fig:TwoLensAndTunable}
\end{center}
\end{figure}

Scan enhancement using two lenses is achieved with a converging and a diverging lens. Such device is depicted in Fig. \ref{fig:TwoLensAndTunable}(a). The ray transfer matrix of this setup is
\begin{equation} \label{eq:M2}
M_{2\mhyphen lens}=
\begin{bmatrix}
1-\frac{d_l}{f_c} & d \\
\frac{1}{f_c f_d}(d_l-f_c-f_d) & 1-\frac{d_l}{f_d}
\end{bmatrix},
\end{equation}
where $d_l$ is the distance between the lenses, and $f_{c/d}$ is the focal length of the converging/diverging lens. For scan enhancement to occur, the parameters must be chosen such that
\begin{equation}
d_l-f_c-f_d=0.
\end{equation}
This condition leads to the desired scan enhancement $\tan\theta=\alpha_2\tan\theta'$, with the angular scan enhancement factor for the two-lens setup defined as
\begin{equation}
\alpha_2=1-\frac{d_l}{f_d}.
\end{equation}

Imagine a linearly phased array located some distance $d$ before the converging lens. The rays of interest now are the parallel rays which would produce a lobe peak at the angle $\theta'$ in the absence of the lenses. The lenses enhance this angle, with the rays leaving the optical device still parallel. Because the enhancement occurs with a linearly phased array no distributed grating lobes are excited by the outer array elements. 

Directivity is again affected. Considering the two outermost rays, the ratio between the effective aperture length on the output of the diverging lens and one on the input of the converging lens leads to
\begin{equation} \label{eq:DirDegTwo}
\frac{D_2(\theta)}{D_{arr}(\theta')} = \frac{1}{\alpha_2}\cdot\frac{\cos\theta}{\cos\theta'},
\end{equation} 
which near broadside is approximately $1/\alpha_2$.

It is worth mentioning that the assumed non-paraxial ray transfer theory and ideal refraction do not place any limits on $d$ nor $d_l$. In the context of the presented theory these values can be arbitrarily small. In practice however it is a question of how well physical lenses conform to these assumptions. We also note that the scan enhancing property of the two-lens enhancer is not limited to it being in the near field of the source array. The two-lens scan enhancement of $\tan\theta=\alpha_2\tan\theta'$ is valid for any ray passing through the system, be it the parallel rays of the near-field case or the rays in the far-field of the source array. Finally, it is also possible to achieve scan enhancement using two converging lenses as in a Kepler telescope configuration. However, this configuration leads to a physically larger device.

\subsection{Scan Enhancement using a Tunable Metasurface} \label{sec:Tunable}

When one considers scan enhancement using a static metasurface (a metasurface whose parameters, such as the surface phase $\phi(x)$, are constant) one almost immediately arrives at the conclusion that the surface must act as some sort of a lens. In \cite{egorov2020theory} and in this publication we showed that single and two lens enhancers exhibit a factor of $1/\alpha$ in their directivity degradation. This factor is undesirable, and it appears that it is unavoidable with static metasurfaces. 

To overcome the poor directivity performance we propose a discretely tunable metasurface concept to achieve the enhancement (see Fig. \ref{fig:TwoLensAndTunable}(b)). Such a surface would be placed in the radiating near-field of the source array. To achieve a desired scan enhancement the metasurface needs to exhibit only two states -- one where the surface exhibits a linear phase ($\phi(x) = k_s x$) and the second state is the negative of the first ($\phi(x) = -k_s x$).

Given an array which can be scanned up to $\theta_{max}$, its scan range can be increased to $\alpha_t\theta_{max}$ by choosing
\begin{equation} \label{eq:ks}
k_s = k \left( \sin\alpha_t\theta_{max} - \sin\theta_{max} \right),
\end{equation}
which was obtained simply by rearranging the generalized refraction law with the appropriate surface phase (see (\ref{eq:Snells}) for the statement of the law). To obtain a beam in the range $[0,\alpha_t\theta_{max}]$, the surface is in the first state and the source array is phased linearly and according to the generalized refraction law. The second state is used analogously for the negative beam angles. It is easy to deduce that the directivity degradation is
\begin{equation}
\frac{D_t(\alpha_t\theta)}{D_{arr}(\theta)} = \frac{\cos\alpha_t\theta}{\cos\theta},
\end{equation}
which indeed is not reduced by the factor $1/\alpha$ as in (\ref{eq:DirDegSingle}) and (\ref{eq:DirDegTwo}).

We note that a binary (i.e. two-state) tunable metasurface cannot have an arbitrary $\alpha_t$. It is easy to show from (\ref{eq:Snells}) and (\ref{eq:ks}) that $\alpha_t$ is limited to
\begin{equation}
\alpha_t \le \frac{\arcsin(2\sin\theta_{max})}{\theta_{max}}.
\end{equation}
Otherwise, the source array would have to be phased beyond $-\theta_{max}$ (impossible by assumption) to obtain a broadside beam. If a larger $\alpha_t$ is desired, a three-state tunable metasurface is required, with the third state being a constant phase state ($\phi(x)=\mathrm{const.}$).

Design of such tunable metasurfaces is a beyond the scope of this paper but could be achieved using embedded switching elements such as PIN diodes or transistors \cite{xu2019realizaton}. The binary nature of the metasurface may perhaps be employed to obtain simplified physical surface designs of what is proposed in the future.

\begin{figure*}[t!]
	\centering
	\begin{subfigure}{0.3\textwidth}
		\centering
		\includegraphics[width=1\textwidth]{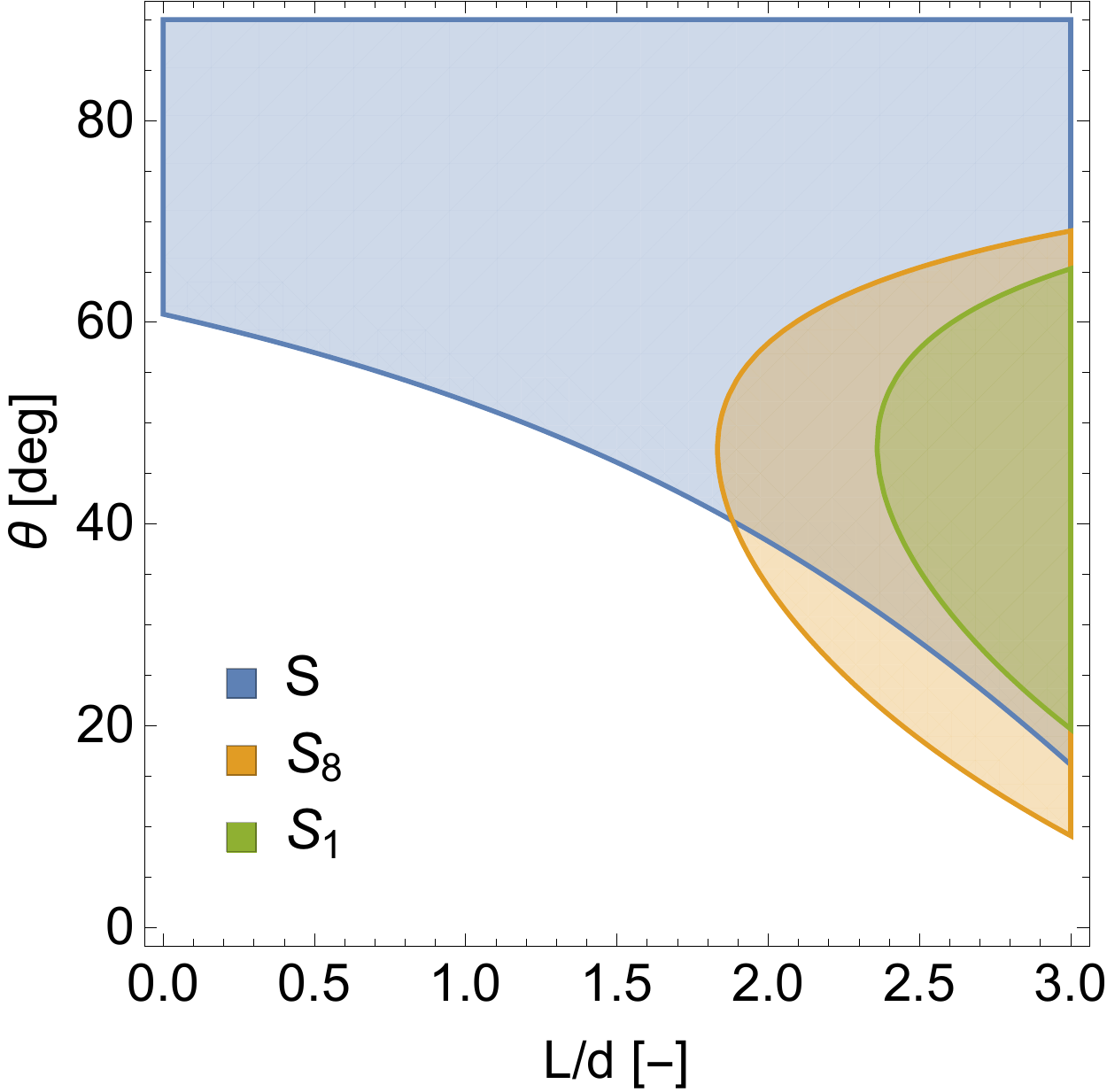}
		\caption{$\alpha{=}2$, $d_e{=}0.6\lambda$}
	\end{subfigure}
	\hfill
	\begin{subfigure}{0.3\textwidth}
		\centering
		\includegraphics[width=1\textwidth]{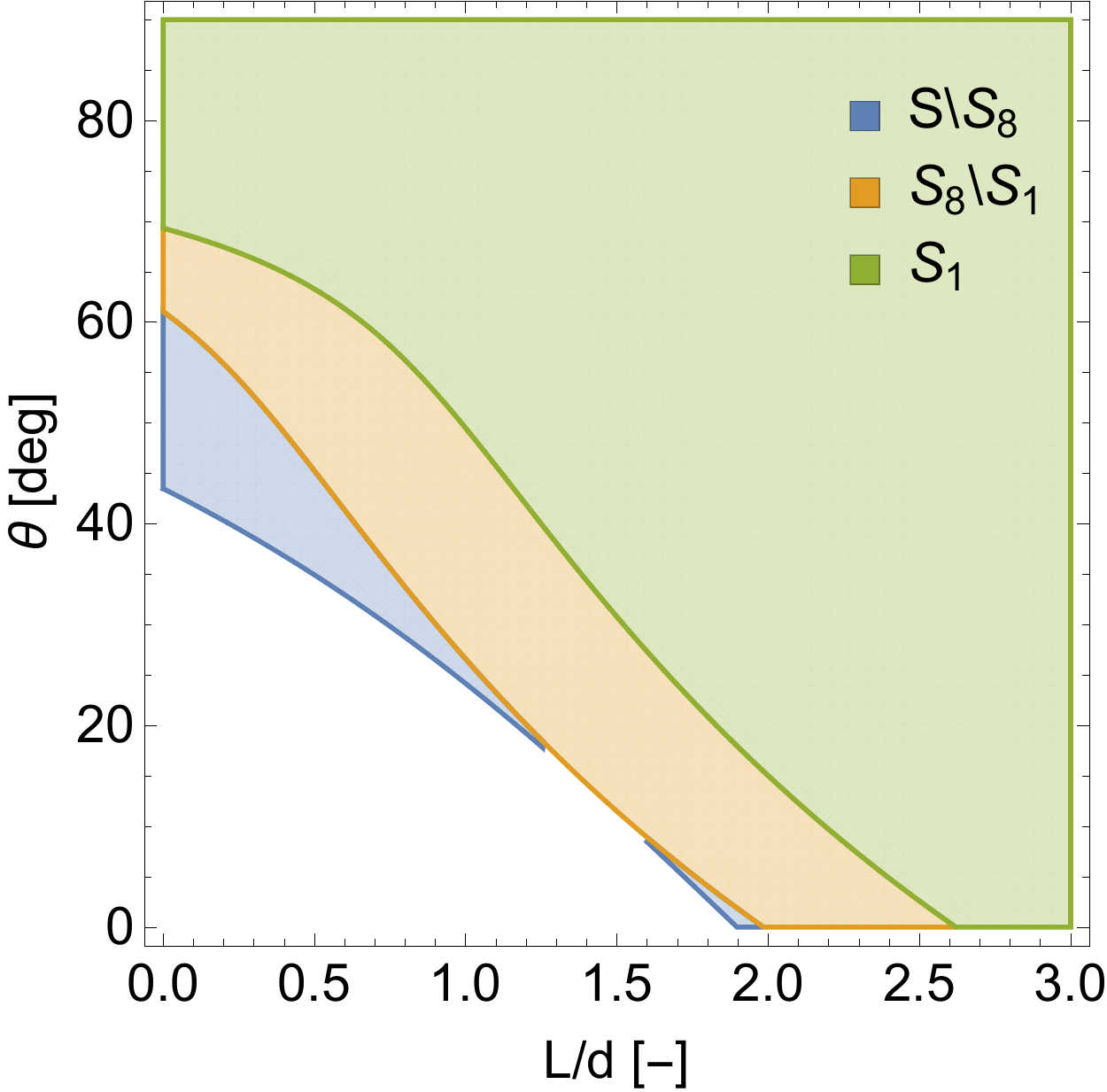}
		\caption{$\alpha{=}2$, $d_e{=}0.7\lambda$}
	\end{subfigure}
	\hfill
	\begin{subfigure}{0.3\textwidth}
		\centering
		\includegraphics[width=1\textwidth]{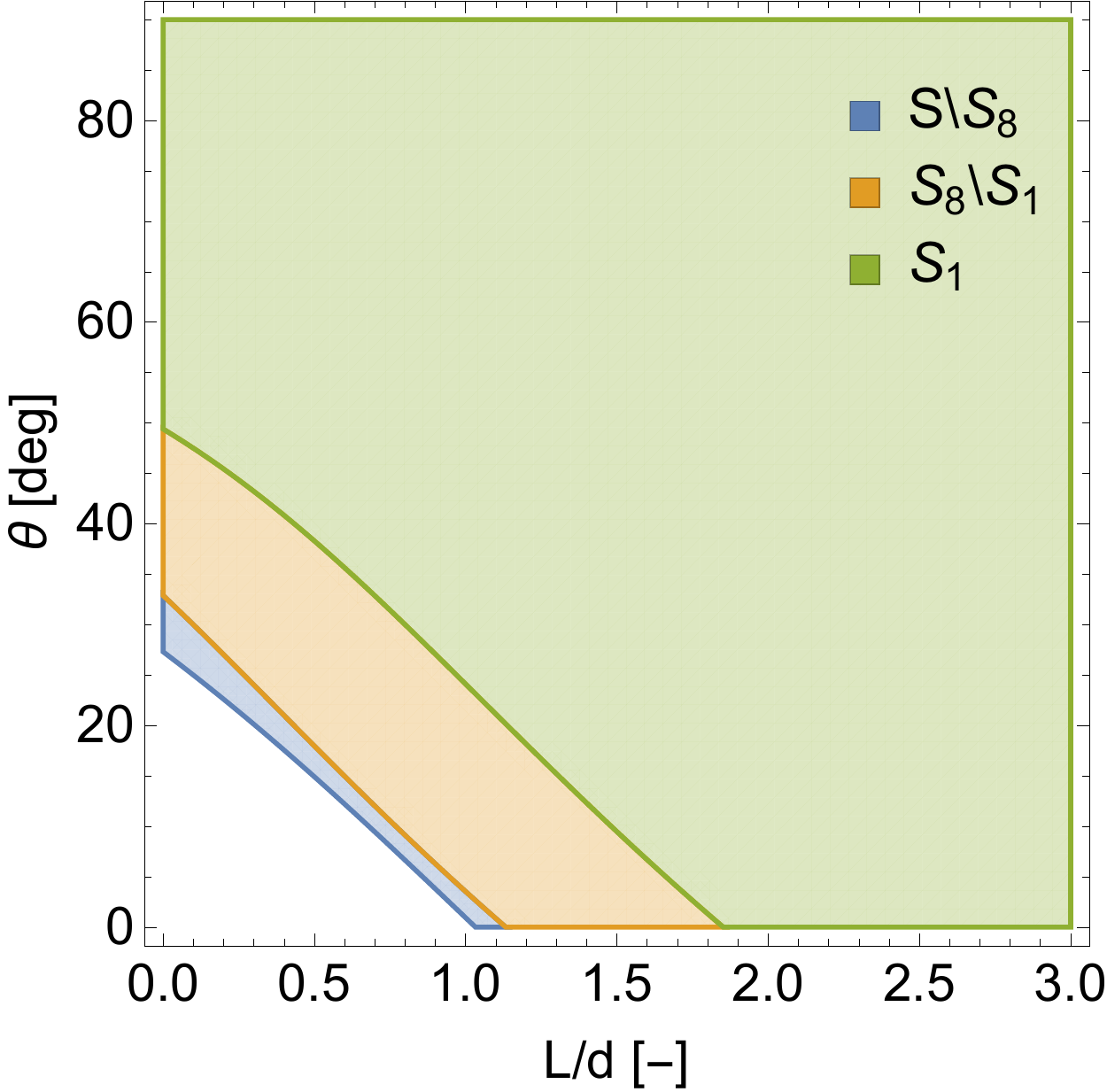}
		\caption{$\alpha{=}2$, $d_e{=}0.8\lambda$}
	\end{subfigure}
	
	\par\bigskip
	
	\begin{subfigure}{0.3\textwidth}
		\centering
		\includegraphics[width=1\textwidth]{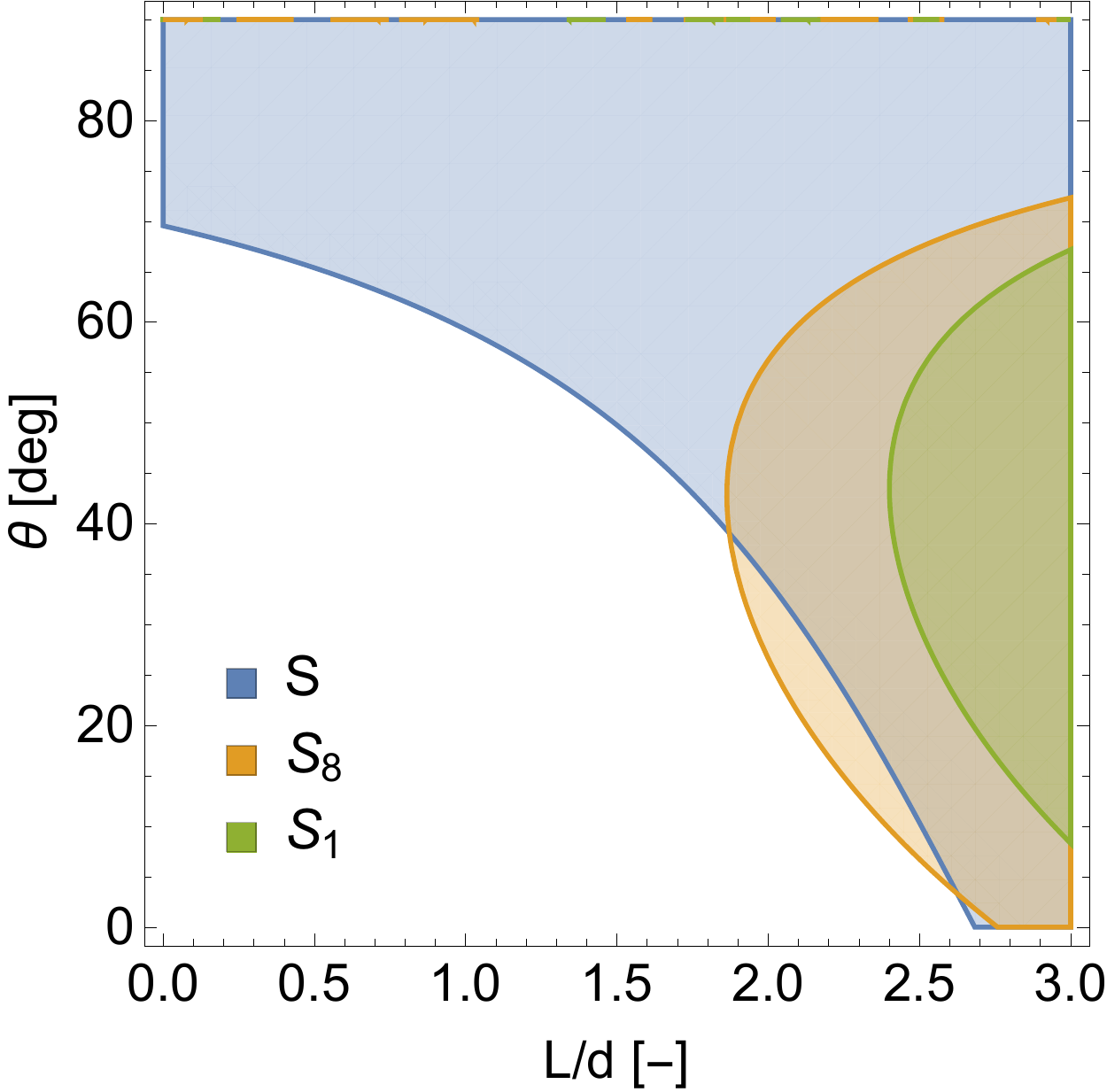}
		\caption{$\alpha{=}3$, $d_e{=}0.6\lambda$}
	\end{subfigure}
	\hfill
	\begin{subfigure}{0.3\textwidth}
		\centering
		\includegraphics[width=1\textwidth]{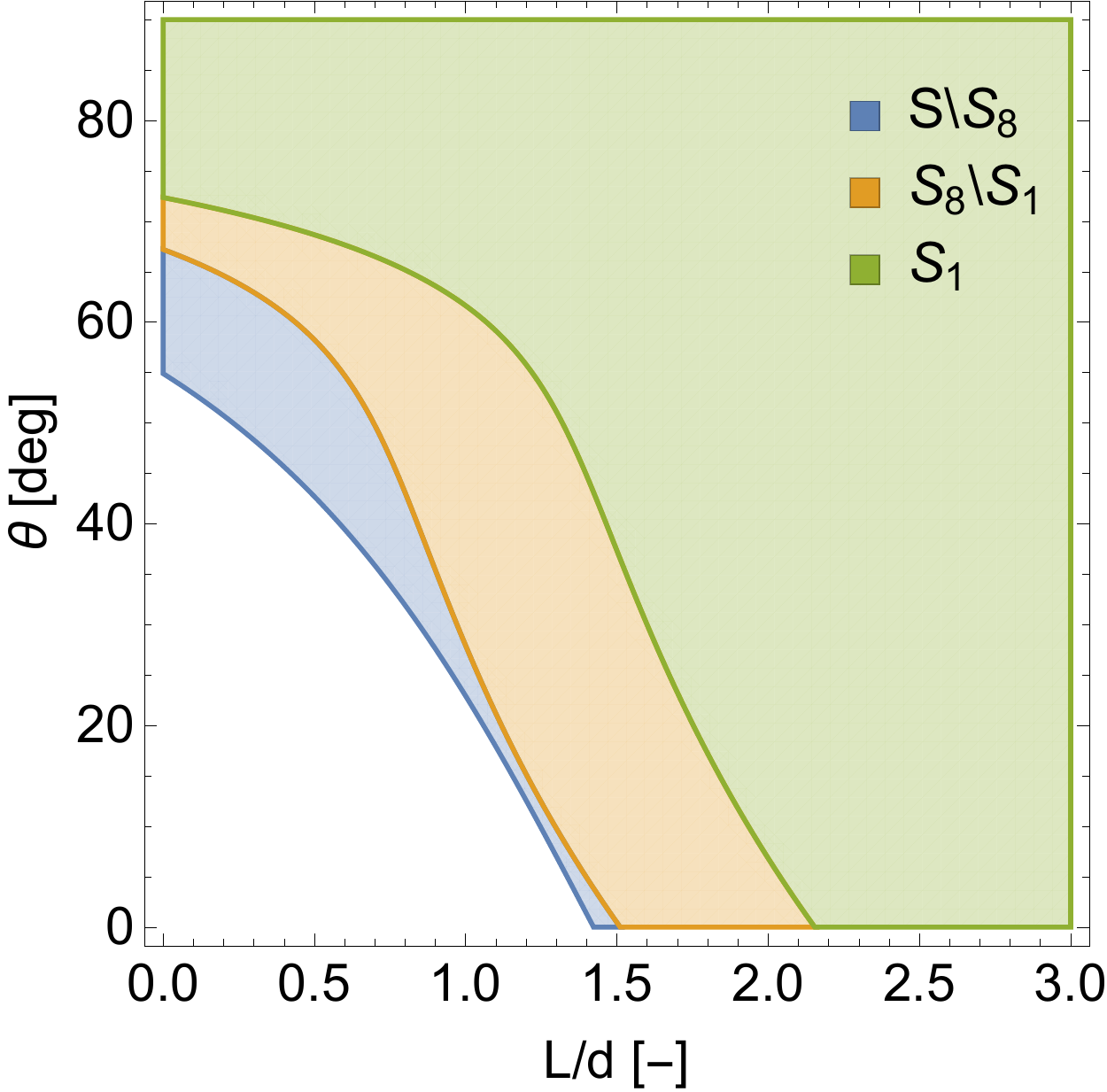}
		\caption{$\alpha{=}3$, $d_e{=}0.7\lambda$}
	\end{subfigure}
	\hfill
	\begin{subfigure}{0.3\textwidth}
		\centering
		\includegraphics[width=1\textwidth]{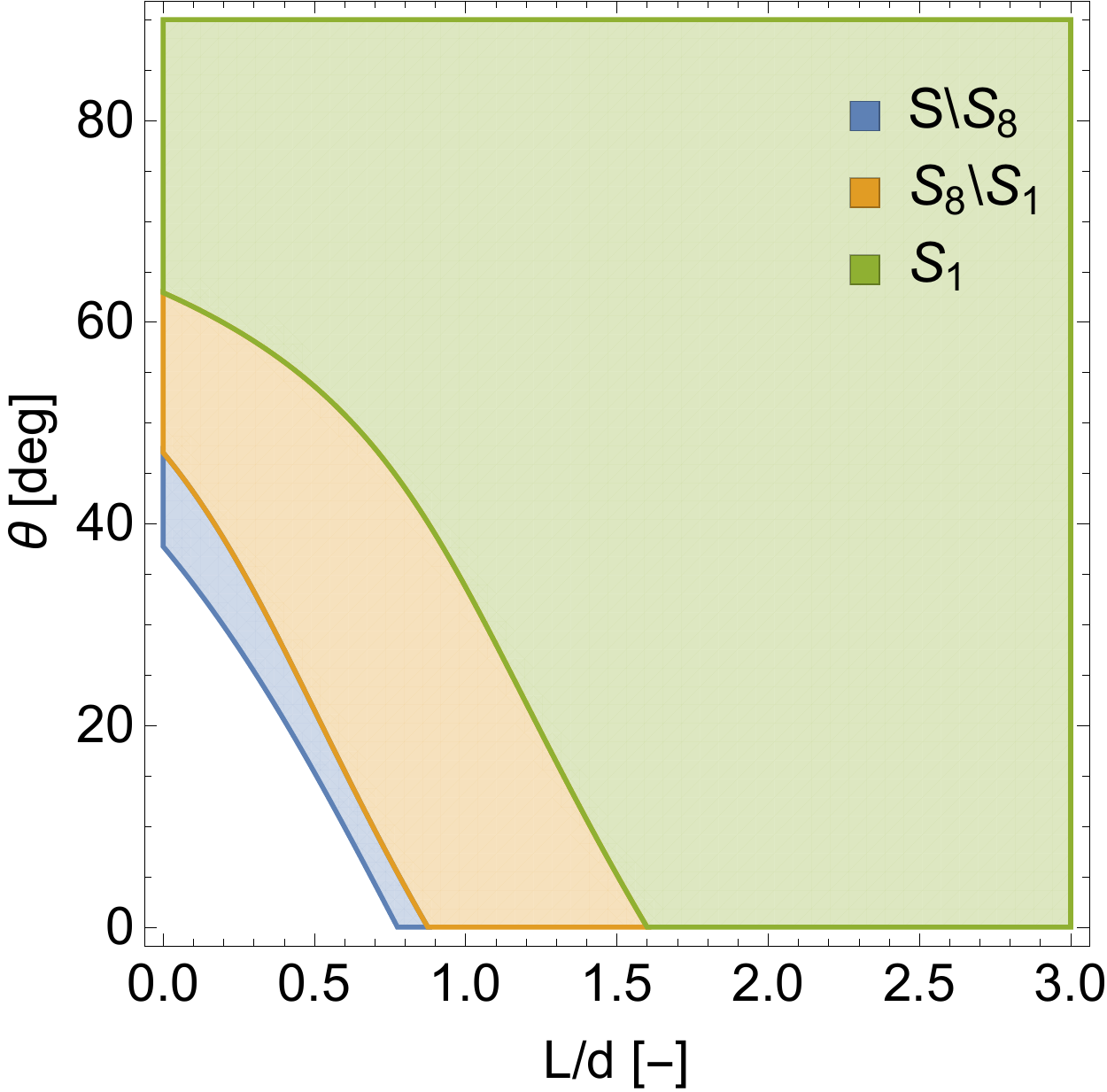}
		\caption{$\alpha{=}3$, $d_e{=}0.8\lambda$}
	\end{subfigure}
	\caption{Behavior of distributed grating lobe inequalities (\ref{eq:GratingLobeCond}) and (\ref{eq:GratingLobeLlim}) for various cases of $\alpha$, $\d_e$ and $d/\lambda$. Note that $A{\setminus}B$ signifies the difference of two sets $A$ and $B$.}\label{fig:DGLIneq}
\end{figure*}

\section{Numerical Studies} \label{sec:NumStudies}

\subsection{Evaluation and Comparison of Distributed Grating Lobe Inequalities} \label{sec:DGLCond}

Let us now study and compare the distributed grating lobe inequalities given by (\ref{eq:GratingLobeCond}) and (\ref{eq:GratingLobeLlim}). Equation (\ref{eq:GratingLobeCond}) is mathematically strict but difficult to interpret, while (\ref{eq:GratingLobeLlim} is less rigorous but descriptive. And so, this comparison is necessary to show that the two expressions are in approximate and qualitative agreement. Note that the inequality of (\ref{eq:GratingLobeLlim}) depends on four independent variables ($L/d$, $\theta$, $\alpha$ and $d_e/\lambda$), while (\ref{eq:GratingLobeCond}) depends on five. This occurs because in (\ref{eq:GratingLobeCond}), $L$ and $d$ cannot be lumped into a single variable as in (\ref{eq:GratingLobeLlim}).

The results of numerical evaluation of (\ref{eq:GratingLobeCond}) and (\ref{eq:GratingLobeLlim}) for various cases are shown in Fig. \ref{fig:DGLIneq}. For a given case of $\alpha$ and $d_e$ let us define $S$ as the set of points for which (\ref{eq:GratingLobeLlim}) is satisfied, and $S_{d/\lambda}$ as the set of points for which (\ref{eq:GratingLobeCond}) is satisfied (with the additional independent variable $d/\lambda$). The plots in Fig. \ref{fig:DGLIneq} depict $S$, $S_1$ and $S_8$. It is immediately apparent that the inequalities are rather different -- (\ref{eq:GratingLobeCond}) and (\ref{eq:GratingLobeLlim}) exhibit different bounding curves and no simple statements can be made with regard to membership properties of the sets. 

Furthermore, Figs. \ref{fig:DGLIneq}a and \ref{fig:DGLIneq}d show some unexpected behavior of the $S_{d/\lambda}$ sets. Consider the behavior of the $S_8$ set of Fig. \ref{fig:DGLIneq}a at $L/d=2.4$. At $\theta \approx 20^\circ$ the associated grating lobe inequality becomes true. It is intuitive to think that at larger angles the phased array would have to be phased in a ``steeper" fashion, i.e. with larger phase differences as $\theta$ increases. This reasoning then suggests that once the inequality becomes true, it remains so at larger $\theta$ values. This however is not the case, since the $S_8$ set does not extend beyond $\theta \approx 65^\circ$. This conceptual inconsistency is easily resolved by realizing that the presented theory is not valid at large values of $L/d$ and $\theta$. The reason for this is the aforementioned fact that non-paraxial ray theory is inconsistent with the generalized law of refraction and it leads to invalid array phasing for large $L/d$ and $\theta$. In Appendix \ref{sec:AppNonParaRerf} we show that for the $\alpha = 2$, $d = 8$ case the presented theory is approximately valid for $L/d \approx 2.4$ up to $\theta \approx 43^\circ$. Beyond $43^\circ$, the prescribed phasing does not lead to the desired outgoing beam and the inequalities become meaningless. 

Let us now consider the range of values of $L/d$ and $\theta$ for which the theory is valid. Table \ref{tbl:ValidRegions} provides approximate ranges for the various cases appearing in Fig. \ref{fig:DGLIneq}. Note that for a given case these ranges are not unique. How these values are obtained is described in more detail in Appendix \ref{sec:AppNonParaRerf}. For large $d_e$ values and within the allowed $L/d$, $\theta$ ranges, it appears that the $S_{\lambda/d}$ sets are contained within the $S$ set. This implies that the inequality of (\ref{eq:GratingLobeLlim}) is a conservative estimate of when the source array leaks power into the distributed grating lobe, and thus can be used on its own to answer whether a given setup will perform without power leakage. At smaller values of $d_e$ (see Figs. \ref{fig:DGLIneq}a and \ref{fig:DGLIneq}d) the $S_{d/\lambda}$ sets are not contained within the $S$ set. Nevertheless, we argue one can use either inequality to judge the performance of the scan enhancer. This is because the inequalities were defined with respect to the appearance of the peak of the grating lobe of the two-element array. This strict mathematical condition does not lead to an abrupt change in device performance. For example, even before the grating lobe peak appears, the two element array may already be losing significant power to undesired directions simply because a large part of the grating lobe is already present. Thus, the inequalities of (\ref{eq:GratingLobeCond}) and (\ref{eq:GratingLobeLlim}) represent approximate ``tipping" points in the behavior of the device.

In summary, the result of these comparisons is that (in the regions where the analytical array excitation of (\ref{eq:PhasingRayTheory}) and (\ref{eq:Ie}) are valid) the two inequalities exhibit similar qualitative behavior and we believe that the less rigorous condition (\ref{eq:GratingLobeLlim}) provides a good estimate on whether the directivity of the device is compromised.

\begin{table}[tbh]
  \begin{center}
    \caption{Valid ranges of $L/d$ and $\theta$}
    \label{tbl:ValidRegions}
    \begin{tabular}{cc|cc}
      \toprule 
      \multicolumn{2}{c|}{Cases} & \multicolumn{2}{c}{Valid Ranges} \\
      $\alpha$ & $d/\lambda$ & $L/d$ $(\le)$ & $\theta$ $(\le)$ \\
      \midrule 
      2 & 1 & 2.4 & $43^\circ$\\
        & 8 & 2.4 & $43^\circ$\\
      3 & 1 & 1.6 & $48^\circ$\\
        & 8 & 1.5 & $48^\circ$\\
      \bottomrule 
    \end{tabular}
  \end{center}
\end{table}

\subsection{Simulation of Single-Lens Near-Field Scan Enhancers} \label{sec:SingleLensSims}

We now study single-lens scan enhancers with full-field simulations using Comsol. The diverging lens is implemented in Comsol via a three-layer ideal impedance structure. The process with which the required impedance values of the three layers can be obtained is discussed in detail in \cite{egorov2020theory}. Here we take the lens designs for granted. All simulated single-lens enhancers of this section are illuminated by 16-element antenna arrays of varying lengths.

\begin{figure}[t!]
	\centering
	\begin{subfigure}{\columnwidth}
		\centering
		\includegraphics[width=3.5in]{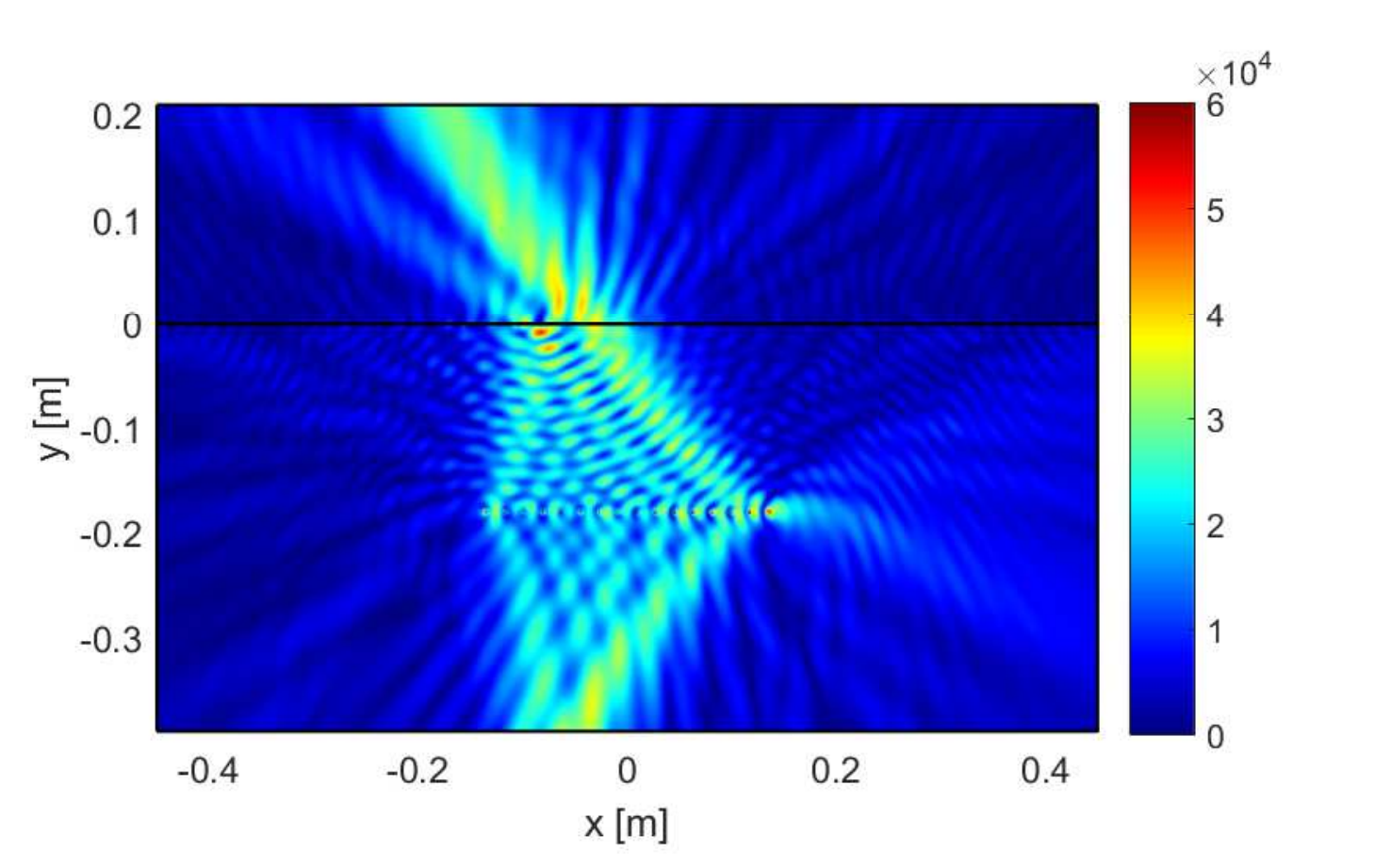}
		\caption{$d_e=0.6\lambda$}
	\end{subfigure}
	
	\par\bigskip
	
	\begin{subfigure}{\columnwidth}
		\centering
		\includegraphics[width=3.5in]{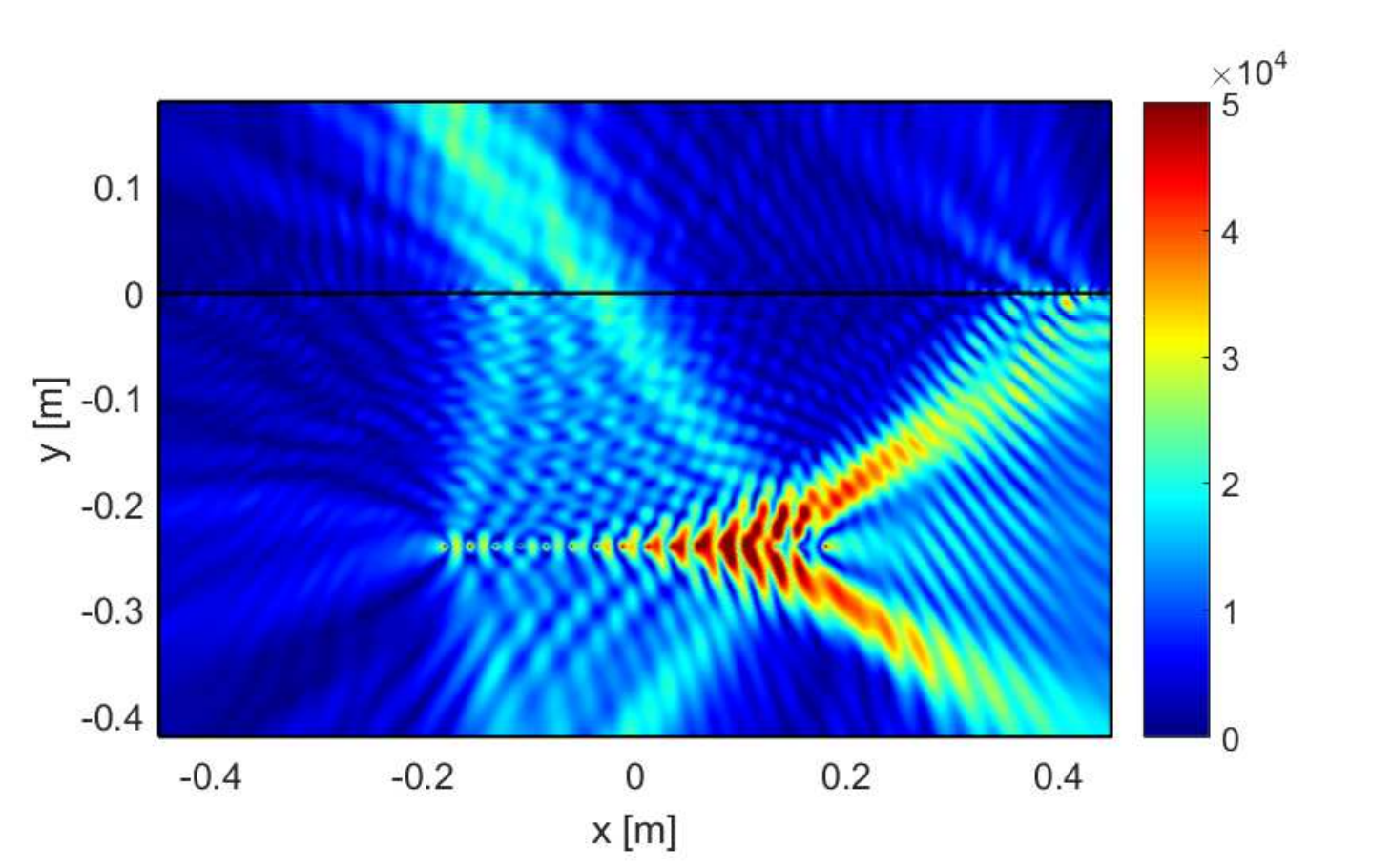}
		\caption{$d_e=0.8\lambda$}
	\end{subfigure}
	\caption{$|\mathbf{E}|$ is depicted. Single-lens angle doublers illuminated by 16-element arrays. Both scenarios have $L/d=1.5$ and desired $\theta$ is 30$^\circ$.}\label{fig:MaciComsolField}
\end{figure}

Fig. \ref{fig:MaciComsolField} shows two angle doublers, both operating with $L/d=1.5$. The two cases (a) $\d_e=0.6\lambda$ and (b) $d_e=0.8\lambda$ were chosen with Fig. \ref{fig:DGLIneq} in mind. In (a) no distributed grating lobe is expected, while (b) should be degraded by it. Indeed this is observed in the figure. In (b) significant power leakage is clearly visible in undesired directions.

\begin{figure}[t!]
	\centering
	\begin{subfigure}{\columnwidth}
		\centering
		\includegraphics[width=3.5in]{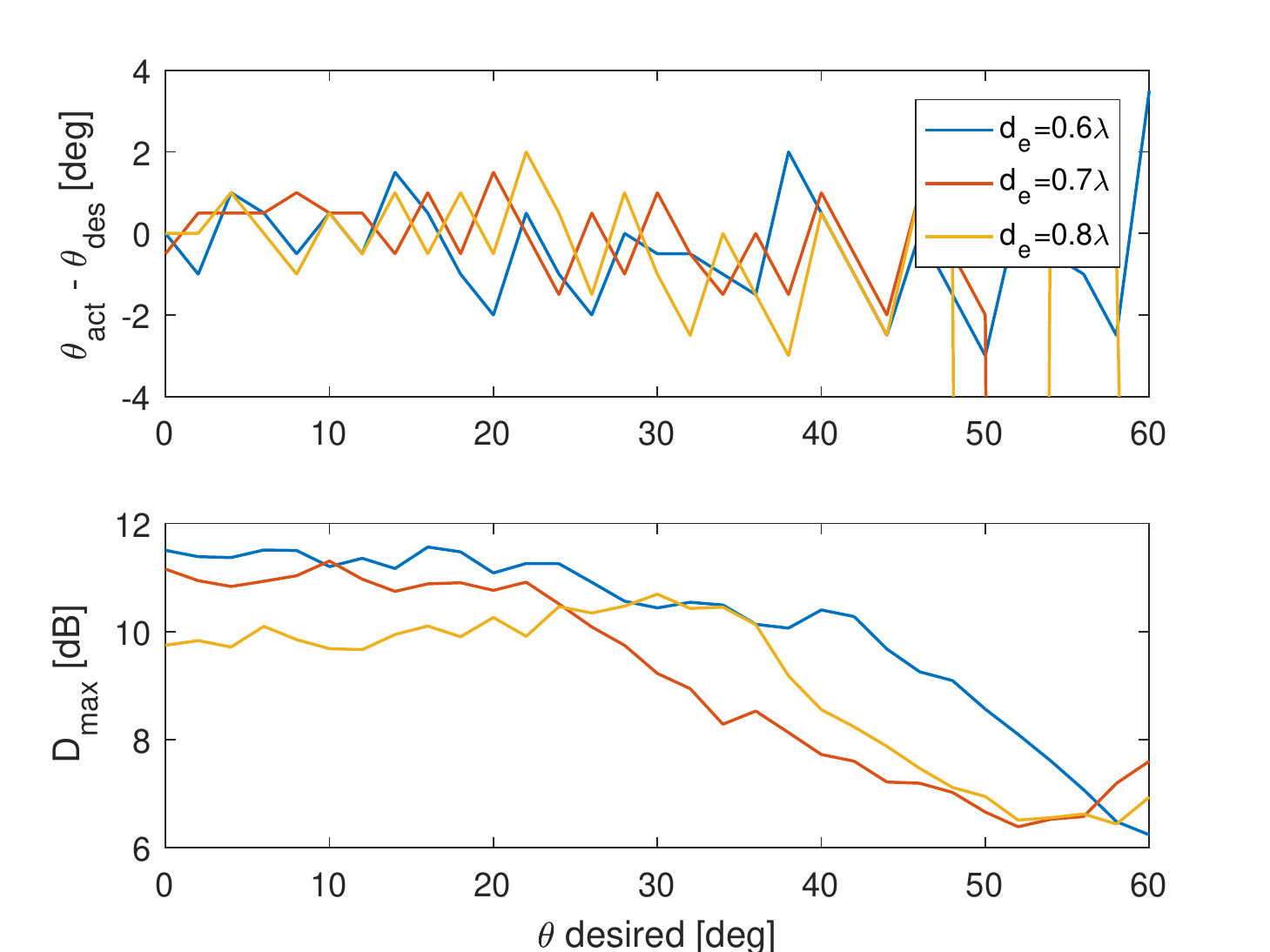}
		\caption{$\alpha{=}2$}
	\end{subfigure}
	
	\par\bigskip
	
	\begin{subfigure}{\columnwidth}
		\centering
		\includegraphics[width=3.5in]{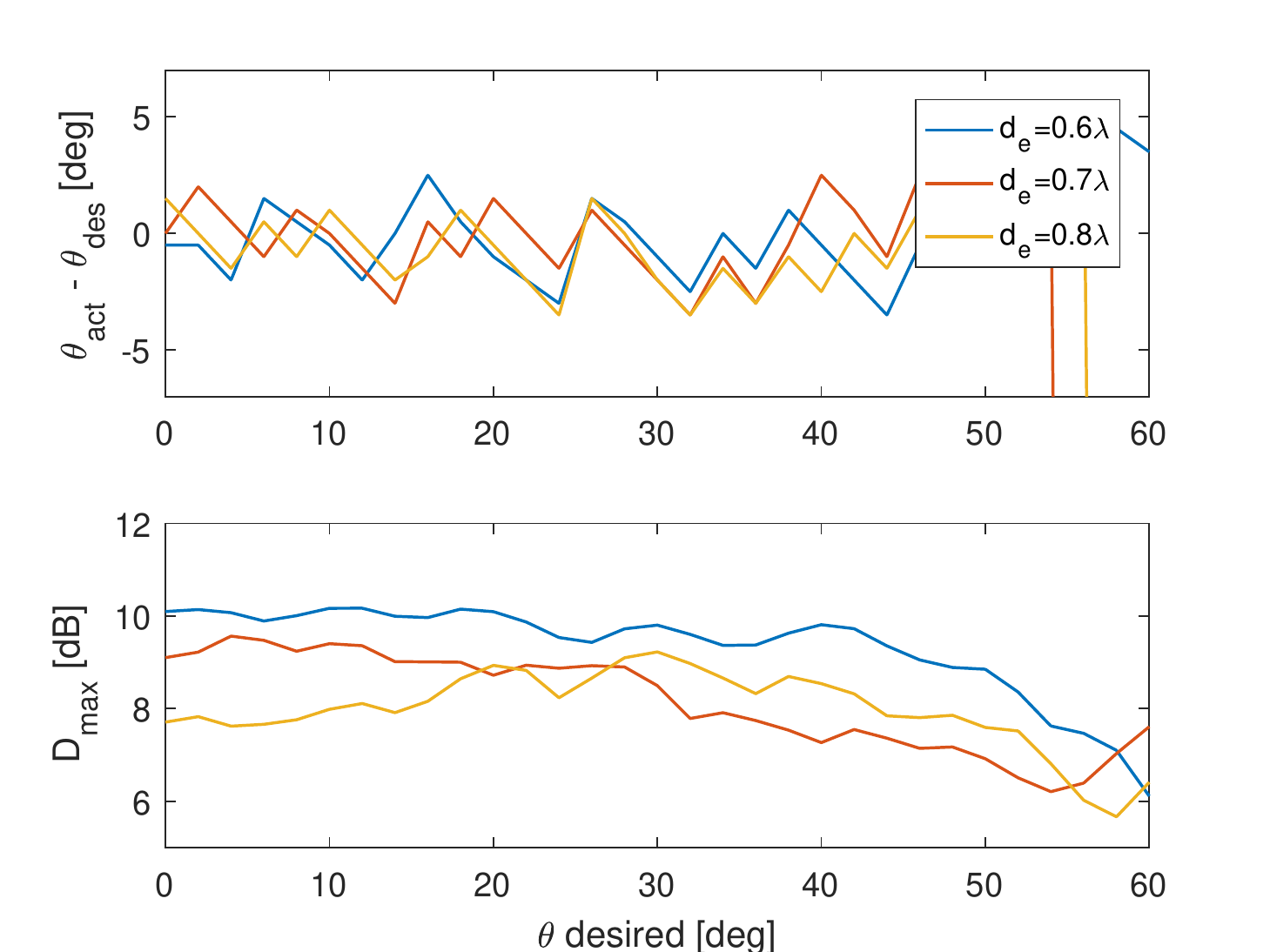}
		\caption{$\alpha{=}3$}
	\end{subfigure}
	\caption{Performance of single-lens enhancers. In all cases a 16-element array is used and $d$ is chosen such that $L/d=1.5$. Note that the peak directivities of the source arrays are 14.7dB ($d_e=0.6\lambda, L=9\lambda$), 15.3dB ($d_e=0.7\lambda, L=10.5\lambda$) and 15.7dB ($d_e=0.8\lambda, L=12\lambda$). (a) Shows the scanning error and directivity performance of angle doublers ($\alpha=2$) corresponding to the three cases shown in Fig. \ref{fig:DGLIneq}(a)-(c). (b) Shows the scanning error and directivity performance of angle triplers ($\alpha=3$) corresponding to the three cases shown in Fig. \ref{fig:DGLIneq}(d)-(f).}\label{fig:MaciSim}
\end{figure}

Fig. \ref{fig:MaciSim} depicts the performance of various single-lens scan enhancers. Six cases were simulated, corresponding to the six cases of Fig. \ref{fig:DGLIneq}. In each simulated case, $d$ was chosen such that $L/d=1.5$.

Fig. \ref{fig:MaciSim}(a) shows the performance of the angle doublers ($\alpha=2$) at various desired beam angles ($\theta$ desired, or $\theta_{des}$). The top subplot shows the scan error of the device, given by the difference between the actual angle at which the peak directivity of the beam is obtained and the desired angle ($\theta_{act}-\theta_{des}$). The scan performance of angle doublers is accurate, and is within $\pm 2^\circ$ of the desired direction for most of the simulated range. It can be seen that near $50^\circ$, the scan errors of the $d_e=0.7\lambda$ and $d_e=0.8\lambda$ cases spike, which corresponds to the directivity of the distributed grating lobe overcoming that of the desired beam. The bottom subplot of Fig. \ref{fig:MaciSim}(a) shows the peak directivity obtained by the angle doublers. Note that the peak directivities of feeding source arrays are 14.7dB ($d_e=0.6\lambda, L=9\lambda$), 15.3dB ($d_e=0.7\lambda, L=10.5\lambda$) and 15.7dB ($d_e=0.8\lambda, L=12\lambda$). These values were obtained via Comsol simulations and differ from the values given by (\ref{eq:DUmax}) by approximately 3dB because the simulated array produces two lobes (forward and backward), whereas the equation is valid for a single forward beam. At $\theta_{des}=0$, the curves attain 11.5dB, 11.2dB and 9.7dB for $d_e=0.6\lambda, 0.7\lambda, 0.8\lambda$ respectively. This corresponds to a 3.2dB, 4.1dB and 6dB directivity degradation for each case. This is in agreement with behavior predicted by (\ref{eq:DirDegSingle}) and the distributed grating lobe inequalities given by (\ref{eq:GratingLobeCond}) and (\ref{eq:GratingLobeLlim}). According to (\ref{eq:DirDegSingle}) an angle doubler should exhibit a directivity degradation of 3dB at broadside. Furthermore, looking at Fig. \ref{fig:DGLIneq}(a) we see that at $L/d=1.5$ no inequalities are satisfied up to approximately $50^\circ$, implying the array excitation shouldn't leak much power into undesired directions in this range. Looking at Fig. \ref{fig:DGLIneq}(b) the point (1.5,0) is in the vicinity of where the inequalities become satisfied. Indeed we observe a larger degradation of 4.1dB in simulation. In Fig. \ref{fig:DGLIneq}(c) the point is even further in the region where the distributed grating lobes are created, and again further degradation is observed. For beam angles other than $\theta=0$, the peak dB directivity values should ideally follow $10\log_{10}\left(D_{arr}(\theta)/\alpha\right)$ according to (\ref{eq:DirDegSingle}). This behavior is indeed observed (within 0.5dB) for the $d_e=0.6\lambda$ case up to approximately $43^\circ$, which is in agreement with values of Table \ref{tbl:ValidRegions}. The $d_e=0.6\lambda$ case is the only case which follows the ideal theoretical directivity curve because it is the only considered case which does not excite the distributed grating lobe according to Fig. \ref{fig:DGLIneq}. 

Fig. \ref{fig:MaciSim}(b) shows the same type of information as Fig. \ref{fig:MaciSim}(a), but for angle triplers ($\alpha=3$). Again, the three cases of Fig. \ref{fig:DGLIneq}(e)-(f) are considered, with $d$ set such that $L/d=1.5$. Note that the source arrays are the same as in Fig. \ref{fig:MaciSim}(a). Again, the scan error over most of the range is low, within approximately $5^\circ$ for most of the range. Again spikes in scan error are present when the peak directivity of the distributed grating lobe overtakes that of the desired beam. According to (\ref{eq:DirDegSingle}) and using source array directivities from above, at $\theta_{des}=0$ the directivity of the three cases $d_e=0.6\lambda,0.7\lambda,0.8\lambda$ should be 9.9dB, 10.5dB and 10.9dB respectively. The simulated values are 10.1dB, 9.1dB and 7.7dB, which corresponds to 4.6dB, 6.2dB and 8.0dB directivity degradation compared to the source arrays. Once more, simulated directivity performance follows the predictions of the distributed grating lobe inequalities of (\ref{eq:GratingLobeCond}) and (\ref{eq:GratingLobeLlim}). For $\theta_{des}$ other than 0, consider the behavior of the $d_e=0.6\lambda$ case. Again, this case is chosen because it is the only one which does not excite the distributed grating lobe according to Fig. \ref{fig:DGLIneq}. The directivity curve of this case follows the theoretical $10\log_{10}\left(D_{arr}(\theta)/\alpha\right)$ accurately for most of the operating region.

In summary, the near-field single-lens scan enhancers can perform extremely well with regard to scan and theoretical directivity performance -- as long as the distributed grating lobes are not excited.

\subsection{Simulation of Two-Lens Near-Field Scan Enhancers} \label{sec:TwoLensSims}

Let us now simulate the proposed two-lens enhancers, again with full-field simulations in Comsol. Two three-layer impedance structures are needed now for simulation -- one making a converging and another a diverging lens. Although \cite{egorov2020theory} only discusses the ``design" of diverging lenses in detail, the converging lenses can be easily obtained by the same arguments and we will not discuss the implementation in detail. 

\begin{figure}[t!]
\begin{center}
\includegraphics[width=3.5in]{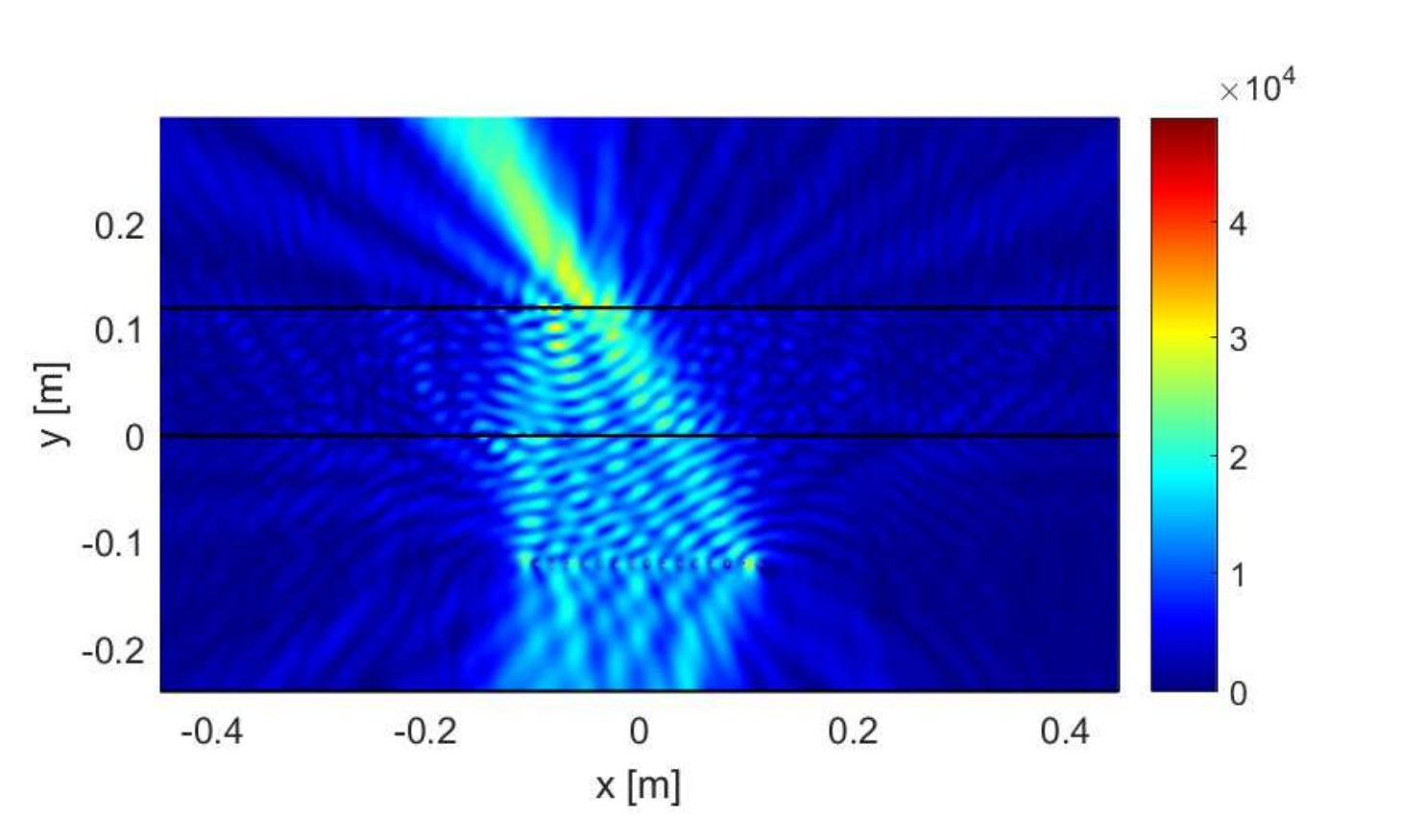}
\caption{$|\mathbf{E}|$ is depicted. A two-lens angle doubler ($\alpha_2=2$) enhancing a 15$^\circ$ incident beam.}\label{fig:TwoLensComsolField}
\end{center}
\end{figure}

Fig. \ref{fig:TwoLensComsolField} shows an angle doubler enhancing an incident beam, with the magnitude of the electric field shown. Now the source array is a 16-element, $\lambda/2$-spaced array. The shown scenario has $d,d_l=4\lambda$. The array is phased to 15$^\circ$ off-broadside and indeed the two lenses steer the beam towards approximately 30$^\circ$.

\begin{figure}[t!]
	\centering
	\begin{subfigure}{\columnwidth}
		\centering
		\includegraphics[width=3.5in]{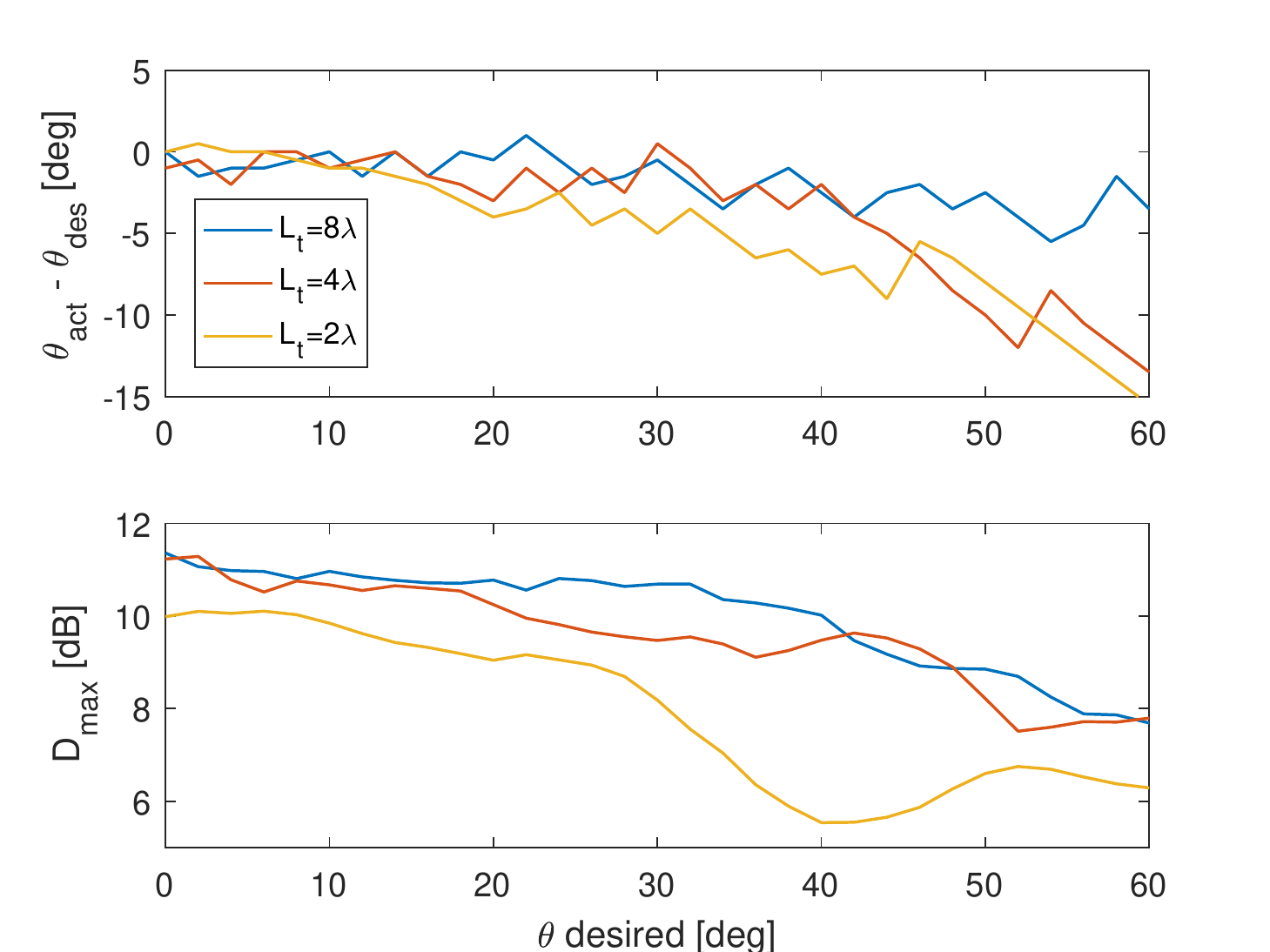}
		\caption{$\alpha_2{=}2$}
	\end{subfigure}
	
	\par\bigskip
	
	\begin{subfigure}{\columnwidth}
		\centering
		\includegraphics[width=3.5in]{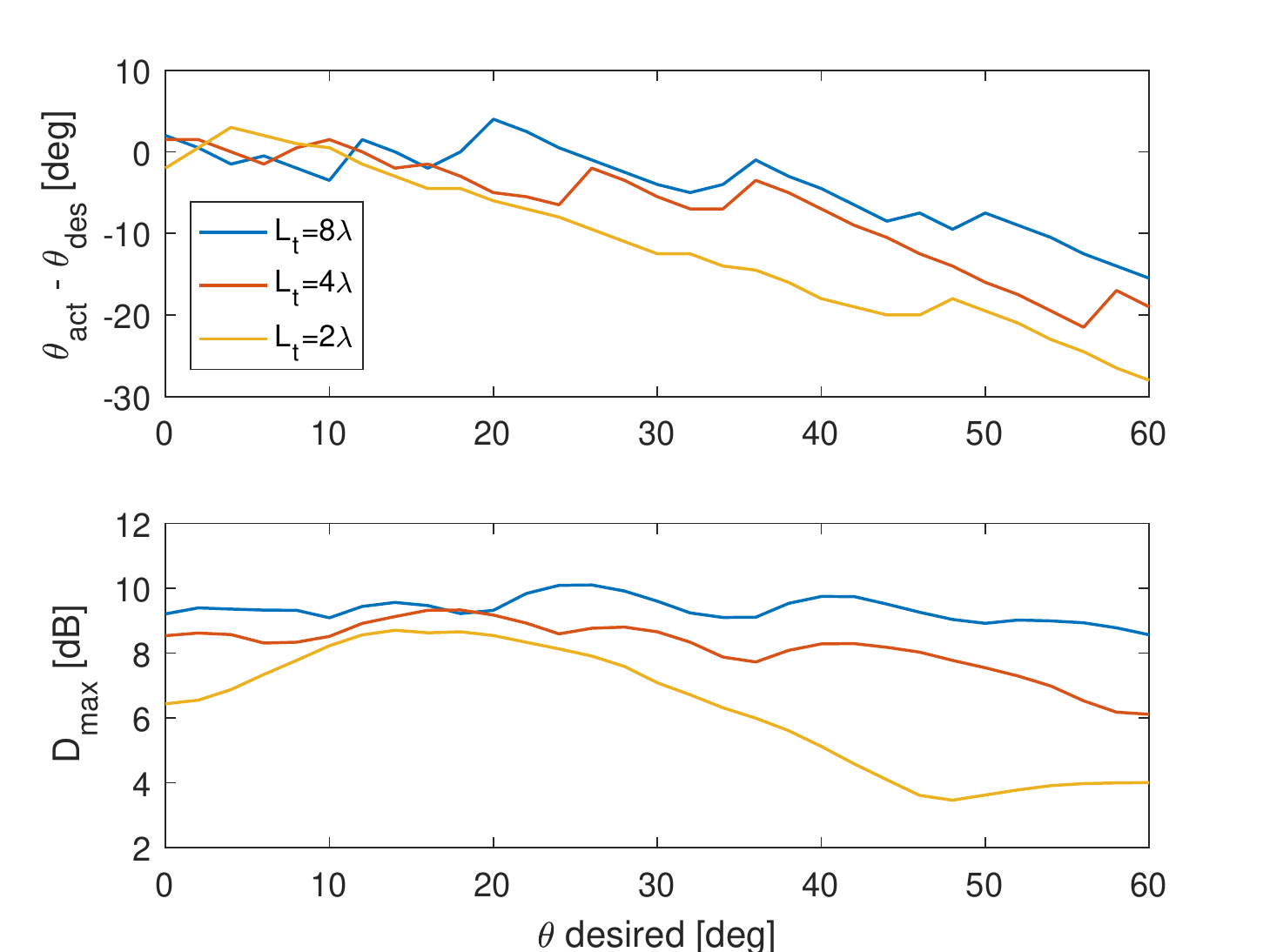}
		\caption{$\alpha_2{=}3$}
	\end{subfigure}
	\caption{Performance of two-lens enhancers. In all cases a 16-element, $\lambda/2$-spaced array is used. The source array exhibits a peak directivity of 13.9dB. Different enhancers, corresponding to various overall device lengths ($L_t=d+d_l$, with $d=d_l$), were simulated. (a) Shows the scanning error and directivity performance of angle doublers ($\alpha_2=2$). (b) Shows the scanning error and directivity performance of angle triplers ($\alpha_2=3$).}\label{fig:TwoLensSim}
\end{figure}

Fig. \ref{fig:TwoLensSim} shows the simulated behavior of various two-lens scan enhancers. The source array on its own has peak directivity of 13.9dB. The overall size of the enhancing devices along the optic axis ($L_t=d+d_l$) is varied, and geometric parameters are chosen such that $d=d_l$. Note that the data is plotted versus desired beam angle $\theta_{des}$, while the array itself is scanned to $\theta_{des}/\alpha_2$. Looking at the scan performance of angle doublers and triplers, it is clearly visible that making $L_t$ too small leads to increased scan errors at large values of $\theta_{des}$. This is attributed to the non-ideality of the simulated ``physical" lenses with respect to non-designed incidence. We note that the scan error of the $L_t=8\lambda$ doubler case stays within $\pm5^\circ$ of 0$^\circ$ for the entire simulated range. For the $L_t=8\lambda$ tripler, the scan error remains within $\pm5^\circ$ up to 40$^\circ$. For directivity performance, at broadside incidence all doublers and triplers expect a 3dB and 4.8dB directivity degradation respectively, which is indeed observed for most of the simulated cases. The $L_t=8\lambda$ doubler approximately follows the theoretical $10\log_{10}\left(D_{arr}(\theta)/\alpha_2\right)$ prediction. The directivity performance of the $L_t=8\lambda$ tripler is also favorable. Furthermore, even the $L_t=4\lambda$ doublers perform well in terms of both scan error and directivity for most of the simulated range of $\theta_{des}$.

\begin{figure}[t!]
	\centering
	\begin{subfigure}{\columnwidth}
		\centering
		\includegraphics[width=3.5in]{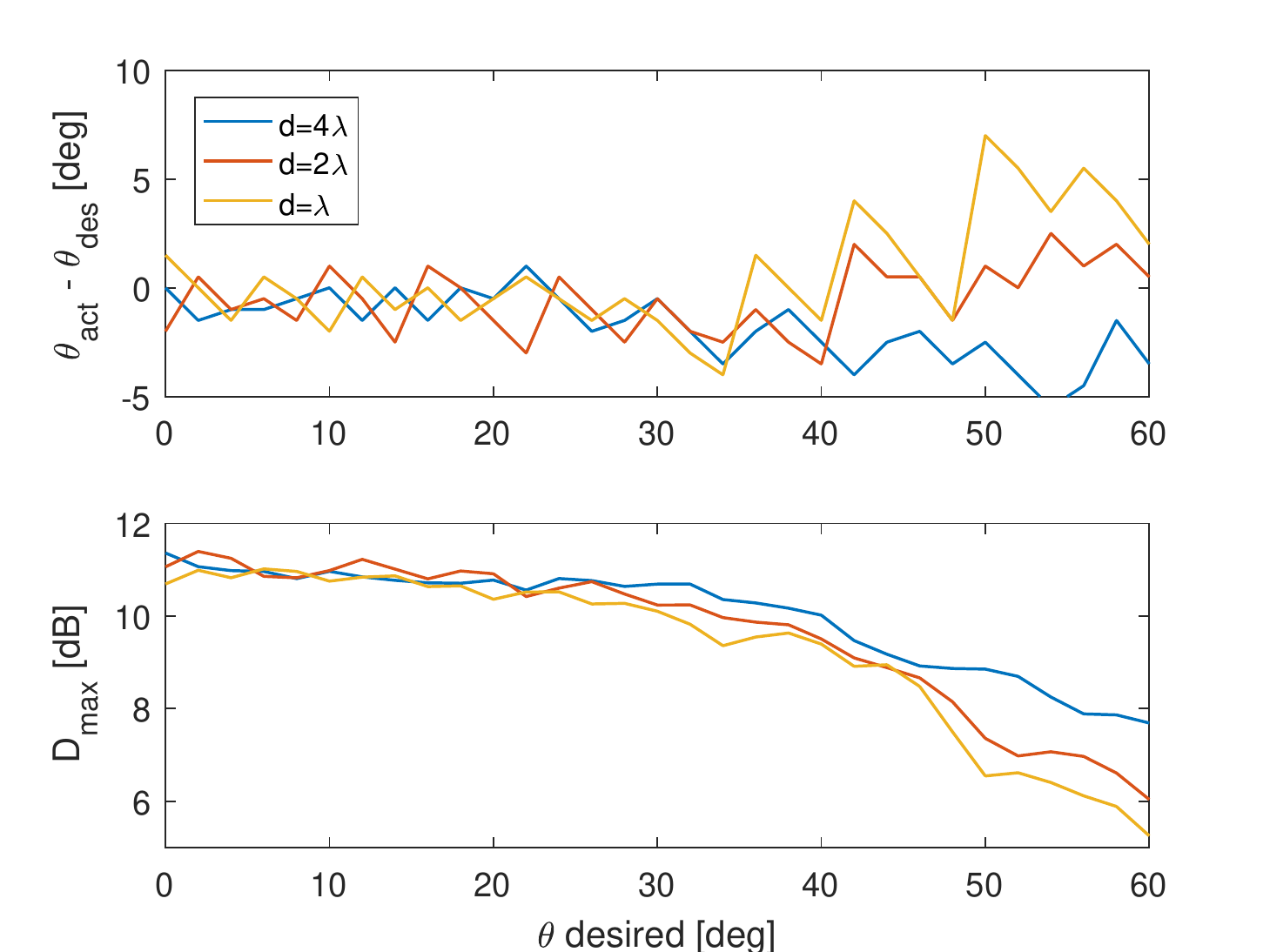}
		\caption{$\alpha_2{=}2$}
	\end{subfigure}
	
	\par\bigskip
	
	\begin{subfigure}{\columnwidth}
		\centering
		\includegraphics[width=3.5in]{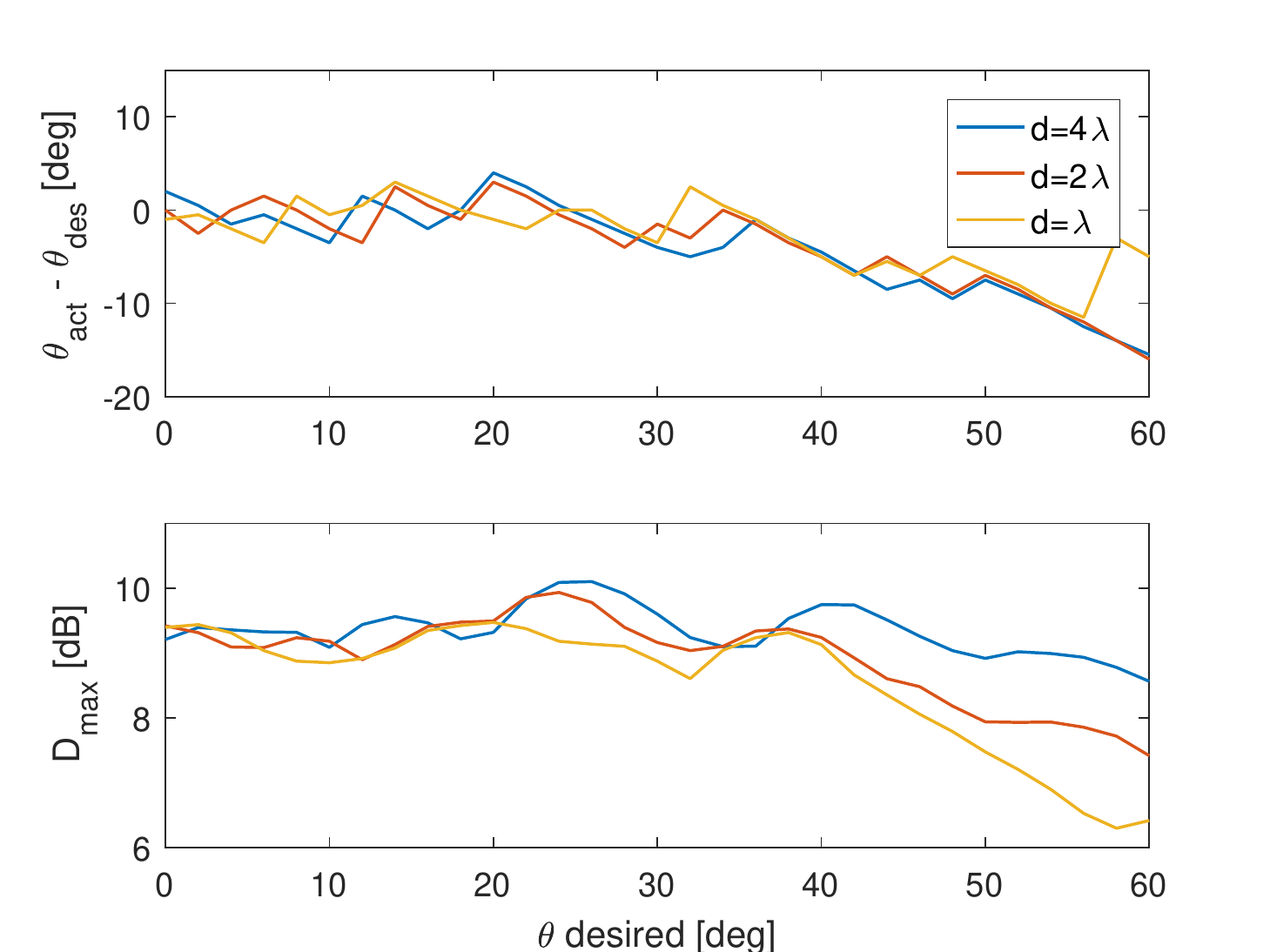}
		\caption{$\alpha_2{=}3$}
	\end{subfigure}
	\caption{Performance of two-lens enhancers. In all cases a 16-element, $\lambda/2$-spaced array is used. The source array exhibits a peak directivity of 13.9dB. In all simulated cases $d_l=4\lambda$. (a) Shows the scanning error and directivity performance of angle doublers ($\alpha_2=2$). (b) Shows the scanning error and directivity performance of angle triplers ($\alpha_2=3$).}\label{fig:TwoLensSim2}
\end{figure}

Fig. \ref{fig:TwoLensSim2} shows the simulation results of angle doublers and triplers all having $d_l=4\lambda$, corresponding to the spacing of the best performing devices of Fig. \ref{fig:TwoLensSim}. The plots show the performance as the source array is brought closer. We see that all simulated angle doublers and triplers maintain good scan and directivity performance up to approximately 40$^\circ$. This shows that, as expected from theory, close placement of the source array to the device does not affect device performance significantly. The performance degradation which occurs at large $\theta_{des}$ is again attributed to the non-ideality of the simulated lenses.

\begin{figure}[t!]
\begin{center}
\includegraphics[width=3.5in]{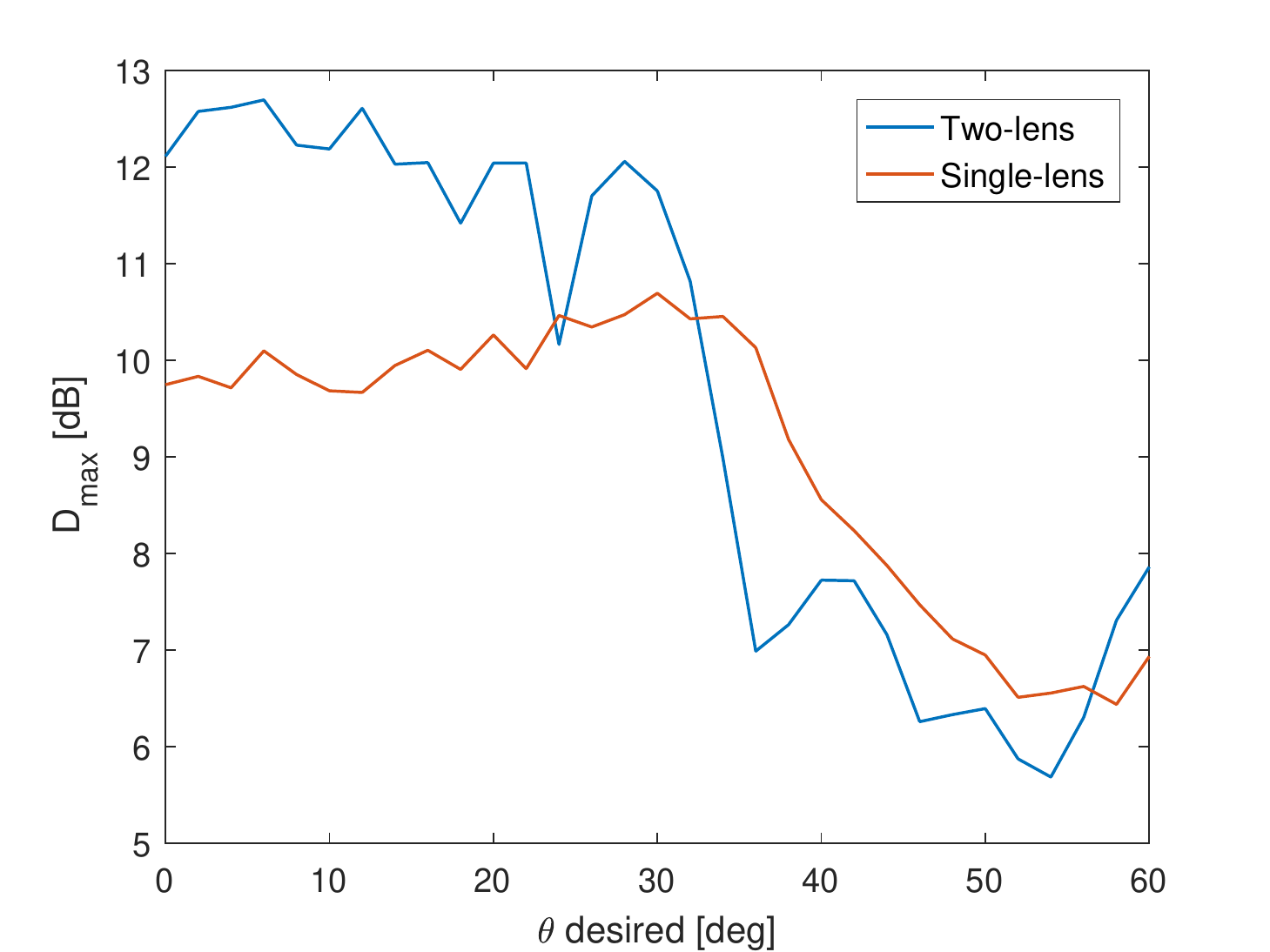}
\caption{Direct comparison of single- and two-lens enhancers. Both devices illuminated by a 16-element, $0.8\lambda$-spaced array. The single-lens device has $d=8\lambda$, while the two-lens device has $d=d_l=4\lambda$.}\label{fig:SingleTwoLensComp}
\end{center}
\end{figure}

So far, the two-lens enhancers were illuminated by a 16-element $\lambda/2$-spaced array. This spacing was chosen such that the array wouldn't produce grating lobes so the behavior of the scan enhancing device itself could be deduced. However, the data provided above does not conclusively show that the two-lens enhancer can provide benefit over the single-lens device. To answer this we simulated a two-lens doubler with $d=d_l=4\lambda$ excited by a 16-element $0.8\lambda$-spaced array and compared the directivity performance to the $d_e=0.8\lambda$ curve of Fig. \ref{fig:MaciSim}a. Fig. \ref{fig:SingleTwoLensComp} shows the directivity performance. In both cases, the overall device length is $8\lambda$. Note that a $0.8\lambda$-spaced array produces the peak of a grating lobe when scanned to about $15^\circ$ off-broadside, and has its directivity compromised even before that. This results in peak directivity obtained for undesired directions for $\theta$ desired from 26$^\circ$ to 34$^\circ$, making the directivity values in this range meaningless. For $\theta \leq 22^\circ$, the directivity of the two-lens device is significantly higher than that of the single-lens device. Near broadside, almost 3dB difference is observed. 

\section{Conclusion}

We have analyzed and proposed a number of scan enhancing devices of phased arrays. The near-field single-lens enhancer, which requires pre-weighting and non-linear phasing of the source array elements for its operation, was studied in detail. Its limitations were uncovered and analyzed. In particular, we found that placing the lens too close to the array leads to a further degradation in directivity performance due to the excitation of a distributed grating lobe. A near-field two-lens enhancer was proposed to achieve enhancement without the distributed grating lobe limitation of the single-lens device. The limitation does not arise for the two-lens device because the source array is excited with a linear phase. A tunable metasurface concept was introduced, which would achieve a scan enhancement without directivity degradation that all other considered devices suffer from. 

Some of the theoretical claims were tested via full-field simulation of proposed devices. The simulation results indeed verified the developed theory. It was found that the analytical source excitation of the single-lens devices indeed provided correct scanning behavior within the acceptable ranges of the theory. The simulation results also showed agreement with the distributed grating lobe limitations of the single-lens device. Simulations of the two-lens devices showed good performance, and finally it was shown that indeed a two-lens device can outperform a single-lens one in terms of the directivity performance for a given scan enhancement.

\appendices

\section{Non-Paraxial Ray Theory \& Generalized Refraction} \label{sec:AppNonParaRerf}

The geometric path of a ray through a lens as dictated by non-paraxial ray theory of Sec. \ref{sec:Prelim} is inconsistent with the generalized refraction law. Non-paraxial ray theory is linear, while the generalized refraction of a lens is not. The linearity of the former theory allowed us to obtain an analytical expression for the element phasing and justify the use of ideal refraction in the time-reversed scenario in Sec. \ref{sec:SingleLensTheory}. 

In accordance with the assumed $\exp(j\omega t)$ dependence and the coordinate system used throughout this publication, the generalized law of refraction can be written as \cite{yu2011light}
\begin{equation} \label{eq:Snells}
\sin\theta - \sin\theta'_{act} = \frac{1}{k}\frac{\d\phi}{\d x},
\end{equation}
where $\theta'_{act}$ signifies the ``actual" angle that a ray attains upon passing through the phase boundary $\phi(x)$ in the time-reversed scenario. We say actual because the generalized refraction law is obtained from the physically fundamental Fermat's principle, while the quotation marks are there because there is no particular reason for a physical metasurface structure to conform to some $\phi(x)$ for all possible incidence scenarios. 

Plugging in the postulated phase of an ideal lens given by (\ref{eq:LensPhase}) we obtain
\begin{equation} \label{eq:LensGRL}
\sin(\theta) - \sin\left(\theta'_{act}\right) = \frac{\sgn(f)x}{\sqrt{x^2+f^2}}.
\end{equation}
With the time-reversed scenario in mind, this equation describes how a ray of interest is refracted by the lens towards the array. The above expression in general leads to different results compared to (\ref{eq:ABCDrelation3}). At the same time, non-paraxial ray theory was defined such that lens refraction would lead to correct behavior of canonical rays. The generalized law of refraction and non-paraxial ray theory indeed agree for canonical rays. For example, consider a case where $\theta = 0$ and $f < 0$, which corresponds to one ``type" of canonical rays refracted by a diverging lens. Equations (\ref{eq:ABCDrelation3}) and (\ref{eq:LensGRL}) reduce to 
\begin{align}
\tan(\theta') &= -\frac{x}{f}, & \sin\left(\theta'_{act}\right) &= \frac{x}{\sqrt{x^2+f^2}}.
\end{align}
The right-hand sides of the above equations refer to the three sides of the same right-angle triangle, and so $\theta' = \theta'_{act}$  for $\theta = 0$. Similar comparisons can be made for all other canonical rays.

The next natural questions are how accurate is the non-paraxial ray theory, and does it offer any benefit over the paraxial approximation ($\theta,\theta',x \approx 0$)? Under the paraxial approximation, time-reversed lens refraction is described by
\begin{equation} \label{eq:Paraxial}
\theta'_{par} = -\frac{x}{f} + \theta.
\end{equation}
It is possible to obtain an analogous analytical expression to (\ref{eq:PhasingRayTheory}), but it is of not much interest here. Note that the above paraxial approximation can be obtained from either the generalized refraction law, or the non-paraxial ray theory. 

\begin{figure}[t!]
\begin{center}
\includegraphics[width=3in]{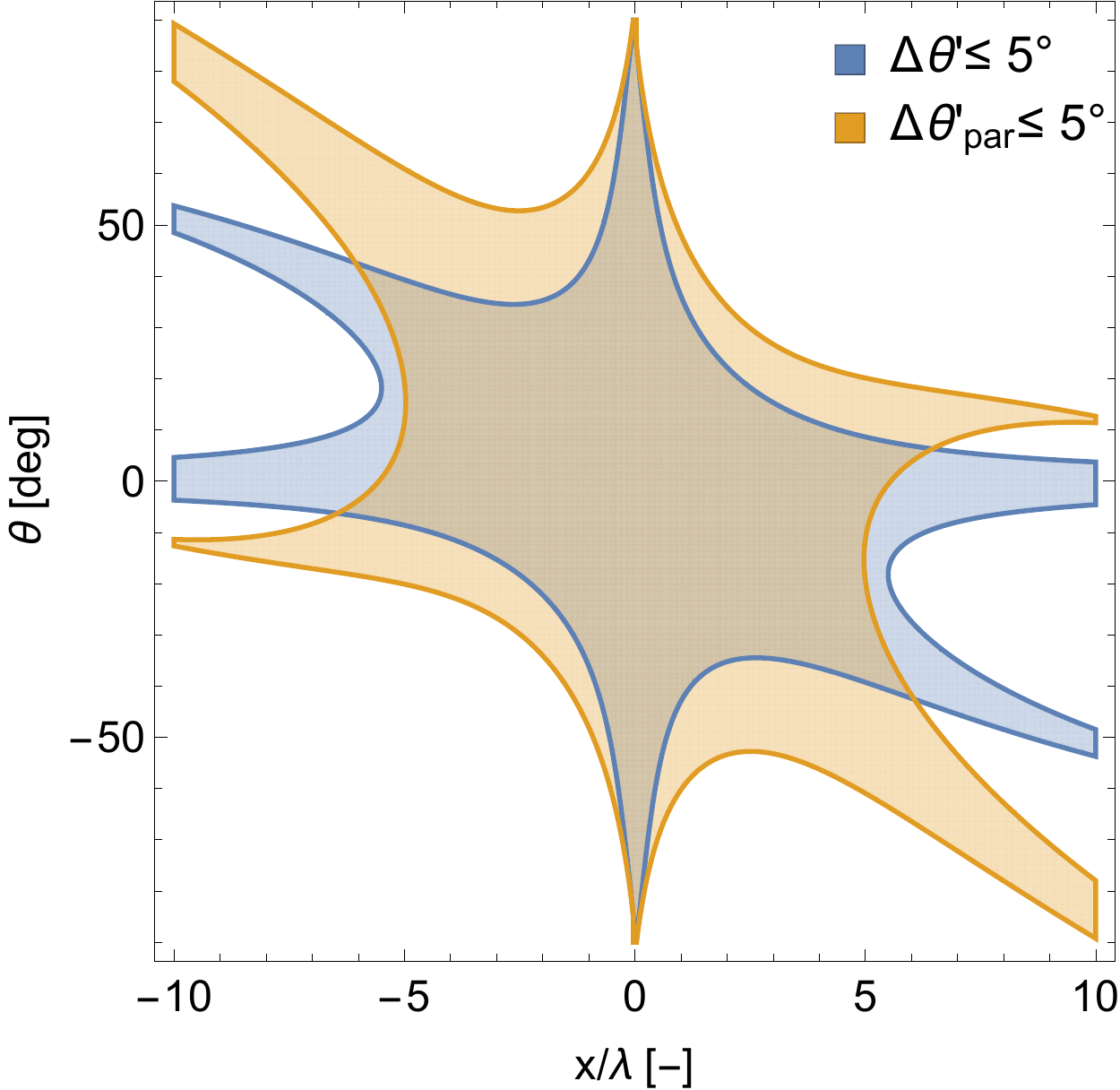}
\caption{Regions of $\theta$ and $x/\lambda$ values for which non-paraxial theory and the paraxial approximation predict refraction accurately (within $5^\circ$ of $\theta'_{act}$). Evaluated for the case of $\alpha = 2$ and $d = 8\lambda$.}\label{fig:ThetaPError}
\end{center}
\end{figure}

To proceed with the comparison, imagine a time-reversed beam traveling at an angle $\theta$ towards the lens. The rays hit the lens at various $x$ points and are refracted. Let us define the following direction errors for any beam ray:
\begin{align}
\Delta\theta'(x,\theta) &= |\theta'(x,\theta)-\theta'_{act}(x,\theta)|, \\
\Delta\theta'_{par}(x,\theta) &= |\theta'_{par}(x,\theta)-\theta'_{act}(x,\theta)|,
\end{align}
where ray directions are explicitly shown as functions of $x$ and $\theta$, and are obtained from (\ref{eq:ABCDrelation3}), (\ref{eq:LensGRL}) and (\ref{eq:Paraxial}). Fig. \ref{fig:ThetaPError} shows the behavior of these direction errors for the case of $\alpha = 2$ and $d = 8\lambda$. The two plotted regions depict where $\Delta\theta', \Delta\theta'_{par} \le 5^\circ$ -- we say that as long as the direction error is less than that, the theory/approximation predicting the refraction can be considered valid. It is clearly visible that near $\theta = 0$ the predicted non-paraxial refraction is valid for large regions of $x$, while the paraxial approximation is not. 

\begin{figure}[t!]
\begin{center}
\includegraphics[width=3in]{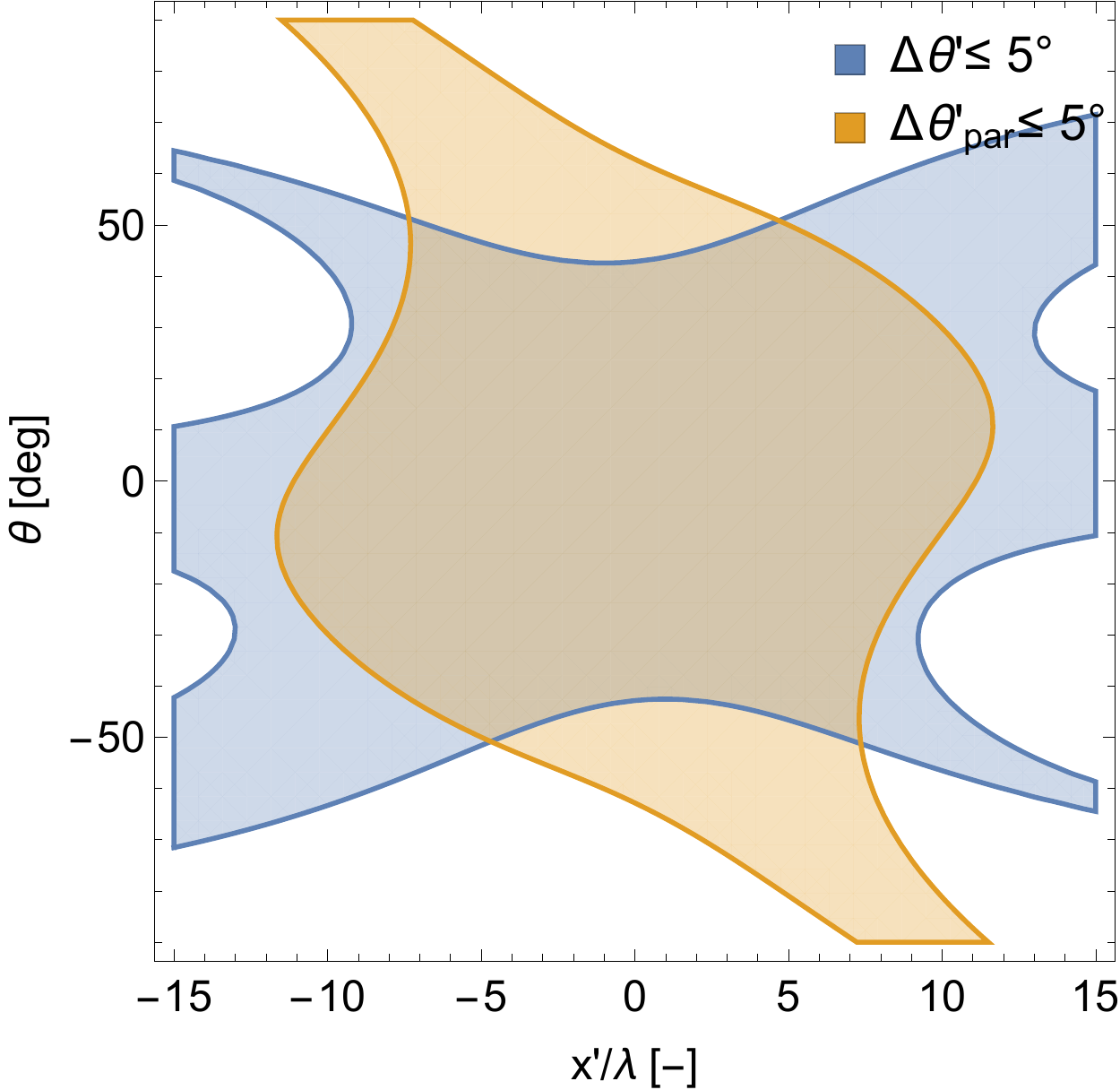}
\caption{Regions of $\theta$ and $x'/\lambda$ values for which non-paraxial theory and the paraxial approximation predict refraction accurately (within $5^\circ$ of $\theta'_{act}$). Evaluated for the case of $\alpha = 2$ and $d = 8\lambda$.}\label{fig:ThetaPErrorxP}
\end{center}
\end{figure}

Looking at Fig. \ref{fig:ThetaPError}, it is not obvious whether non-paraxial theory provides any benefit over the paraxial approximation. Once the time-reversed beam is refracted, its rays propagate towards the array plane and intersect it at some position $x'$. We can express $\theta'$, $\theta'_{act}$ and $\theta'_{par}$ as functions of $x'$ instead of $x$. Refraction as dictated by the non-paraxial ray theory is then given by (\ref{eq:ABCDrelation2}). In the paraxial approximation, the refraction is
\begin{equation}
\theta'_{par} = -\frac{x'}{\alpha f}+\frac{\theta}{\alpha},
\end{equation}
and the generalized refraction law dictates
\begin{equation}
\sin\theta'_{act} - \sin\theta = \frac{x'-d\tan\theta'_{act}}{\sqrt{(x'-d\tan\theta'_{act})^2+f^2}}.
\end{equation}
Note that the above cannot be solved analytically and numerical evaluation had to be employed. In the same way, direction errors are expressed as functions of $x'$, namely $\Delta\theta'(x',\theta)$ and $\Delta\theta'_{par}(x',\theta)$. Fig. \ref{fig:ThetaPErrorxP} plots the regions of $x'$ and $\theta$ where the direction errors are $5^\circ$ or less for the case $\alpha = 2$ and $d = 8\lambda$. Indeed, non-paraxial ray theory offers some benefit over the paraxial approximation -- it is possible to have a larger array while incurring a time-reversed refraction error within $5^\circ$. From the figure it is apparent that a $19\lambda$ wide array can be used to produce refracted beams up to about $43^\circ$ while phased according to non-paraxial ray theory. To achieve the same scan (up to $43^\circ$) with phasing as dictated by the paraxial approximation, the array can be up to $14\lambda$ wide. Note that these limiting values are not unique. For example, phasing as dictated by the non-paraxial ray theory remains valid for an array of width $30\lambda$ up to about $10^\circ$. Note that the values appearing in Table \ref{tbl:ValidRegions} were obtained in a similar fashion by plotting and looking at $\Delta\theta' \le 5^\circ$ and $\Delta\theta'_{par} \le 5^\circ$ for the required cases of $\alpha$ and $d$. We do not provide these plots for the sake of brevity.

\section{Constency of the Far-Field $\rho'(\theta')$ with Maxwell's Equations} \label{sec:AppRhoPrimeValid}

In Sec. \ref{sec:SingleLensTheory} we arrived at a time-reversed refracted ray density of (\ref{eq:rhoprime}), which we assumed expressed far-fields. Let us consider whether the far-fields implied by such $\rho'(\theta')$ indeed satisfy Maxwell's equations. 

The implied far-fields are given by (\ref{eq:Eff}) and (\ref{eq:Hff}). To check the validity of these expressions we assume that the produced electric field is in fact given by (\ref{eq:Eff}). Using (\ref{eq:Eff}) in free-space Maxwell's equations one can calculate the magnetic field, which then with cylindrical coordinates centered around the virtual image point becomes
\begin{equation} \label{eq:Htrue}
\mathbf{H}(r',\theta') = j \frac{e^{-jkr'}}{\omega \mu} \sqrt{\frac{\rho'}{r'}} \left( \frac{1}{2 r' \rho'}\frac{\mathrm{d}\rho'}{\mathrm{d}\theta'}\hat{\mathbf{r}}' + \left( jk + \frac{1}{2r'} \right) \hat{\bm{\theta}}' \right),
\end{equation}
where the $\theta'$ dependence of $\rho'(\theta')$ was omitted for brevity. The $\hat{\mathbf{r}}'$ component and the second term of the $\hat{\bm{\theta}}'$ component correspond to the near-field and go to 0 as $r'$ is increased. The leftover term corresponds to the far-field given by (\ref{eq:Hff}). 

How well does (\ref{eq:Hff}) approximate (\ref{eq:Htrue}) in the region $y \le 0$ (i.e. after lens refraction in the time-reversed scenario)? We say that (\ref{eq:Hff}) is approximately valid (satisfies Maxwell's equations in an approximate fashion) if the two near-field terms in (\ref{eq:Htrue}) are at most 5\% of the far-field term, and so we are interested in the distance $r'_{min}$ at which this condition is satisfied. Note that for moderately sized beams, $1/\rho' \cdot \mathrm{d}\rho'/\mathrm{d}\theta'$ is of order 1. The condition leading to $r'_{min}$ can now be written as
\begin{equation}
\frac{1}{2r'_{min}} = 0.05\cdot\frac{2\pi}{\lambda},
\end{equation} 
which reduces to $r'_{min} \approx 1.6\lambda$. As long as every point in the region $y \le 0$ is further than $1.6\lambda$ from the virtual image, then the assumption that a far-field $\rho'(\theta')$ is established is approximately correct. This in turn limits the focusing power of the enhancing lens to $|f|>1.6\lambda$.

\bibliography{SecondPaper_bibliography}

\bibliographystyle{IEEEtran}

\end{document}